\documentclass[aps,prd,preprintnumbers,superscriptaddress,nofootinbib,notitlepage,floatfix,10pt]{revtex4-2}
\usepackage[pdftex]{graphicx}
\usepackage{bm,latexsym,amsmath,amssymb,amsfonts,mathrsfs}
\usepackage{booktabs} 
\usepackage{color}
\allowdisplaybreaks[1]
\usepackage[pdftex,colorlinks=true,linkcolor=blue,citecolor=cyan,backref=page]{hyperref}
\newcommand*{\D}{\mathrm{d}}
\newcommand*{\mpl}{M_{\mathrm{Pl}}}
\newcommand*{\gn}{G_{\mathrm{N}}}
\newcommand*{\ggw}{G_{\mathrm{GW}}}
\newcommand*{\cgw}{c_{\mathrm{GW}}}
\newcommand*{\ppng}{\gamma^{\mathrm{PPN}}}
\newcommand*{\ppngx}{\gamma^{\mathrm{PPN}}_{\mathcal{X}}}
\newcommand*{\ppnb}{\beta^{\mathrm{PPN}}}
\newcommand*{\ppna}[1]{\alpha^{\mathrm{PPN}}_{#1}}
\newcommand*{\ppnz}[1]{\zeta^{\mathrm{PPN}}_{#1}}
\newcommand*{\ppnxi}{\xi^{\mathrm{PPN}}}
\newcommand*{\TK}{\textcolor[rgb]{1.0,0.1,0.1}}

\begin{document}
\title{PPN meets EFT of dark energy:
Post-Newtonian approximation in higher-order scalar-tensor theories}
%
\author{Jin~Saito}
\email[Email: ]{j\_saito@rikkyo.ac.jp}
\affiliation{Department of Physics, Rikkyo University, Toshima, Tokyo 171-8501, Japan}
\author{Zhibang~Yao}
\email[Email: ]{zhbyao@rikkyo.ac.jp}
\affiliation{Department of Physics, Rikkyo University, Toshima, Tokyo 171-8501, Japan}
\author{Tsutomu~Kobayashi}
\email[Email: ]{tsutomu@rikkyo.ac.jp}
\affiliation{Department of Physics, Rikkyo University, Toshima, Tokyo 171-8501, Japan}
%
\begin{abstract}
We study the post-Newtonian limit of higher-order scalar-tensor theories that are degenerate in the unitary gauge.
They can be conveniently described by the effective field theory (EFT) of dark energy.
We determine all the parametrized post-Newtonian (PPN) parameters in terms of the EFT of dark energy parameters.
Experimental bounds on the PPN parameters are then translated to constraints on the EFT parameters.
We present a Lagrangian of a unitary degenerate higher-order scalar-tensor theory characterized by a single function of the kinetic term of the scalar field whose PPN parameters have the same values as in general relativity.
\end{abstract}
\preprint{RUP-24-3}
\maketitle

\section{Introduction}\label{sec:intro}

A wide variety of modified gravity theories have been proposed so far for tackling the puzzles in the Universe (see, e.g.~\cite{Clifton:2011jh,Heisenberg:2018vsk} for a review).
The basic idea of modifying general relativity (GR) is to introduce, on top of gravitational-wave degrees of freedom (DOFs), new dynamical DOFs that are supposed to resolve the problems in cosmology.
Scalar-tensor theories constitute a well-studied class of modified gravity theories in which the additional DOF is given simply by a scalar field.
There have been significant developments in providing a comprehensive description of scalar-tensor theories since (the rediscovery of) the pioneering work of Horndeski, which exhibited 
the most general scalar-tensor theory in four dimensions having second-order field equations~\cite{Horndeski:1974wa,Deffayet:2011gz,Kobayashi:2011nu}.
To go beyond the Horndeski theory while retaining a single scalar field as a new DOF, one needs to give up the requirement of second-order field equations.
Higher-order field equations result in Ostrogradsky ghost instabilities in general~\cite{Woodard:2006nt}.
However, if the system is degenerate, one can still expect to have healthy theories.
Such ghost-free ``beyond Horndeski'' theories were discovered first by applying an invertible field redefinition to the Horndeski theory~\cite{Zumalacarregui:2013pma} or somewhat in an ad hoc way~\cite{Gleyzes:2014dya} and then were systematically constructed and analyzed under the name of degenerate higher-order scalar-tensor (DHOST) theories~\cite{Langlois:2015cwa,Crisostomi:2016czh,BenAchour:2016fzp,BenAchour:2016cay}
(see~\cite{Langlois:2018dxi,Kobayashi:2019hrl} for a review).
If the gradient of the scalar field is restricted to be timelike and hence it is legitimate to work in the unitary gauge, i.e. to choose the scalar field as a time coordinate, one can further generalize the DHOST theories by weakening the degeneracy conditions so that they are satisfied only in the unitary gauge~\cite{DeFelice:2018ewo}. 
Such a generalized class of healthy higher-order scalar-tensor theories is dubbed the U-DHOST theory.
The extra DOF associated with the would-be Ostrogradsky ghost is not completely gone in the U-DHOST theory.
Nevertheless, the instabilities can indeed be avoided because this extra DOF, dubbed the shadowy mode, obeys an elliptic differential equation on a spacelike hypersurface supplemented with appropriate boundary conditions at infinity and hence does not propagate~\cite{DeFelice:2018ewo, DeFelice:2021hps}.
When working in the unitary gauge, the U-DHOST theory can be mapped into a subclass of the so-called spatially covariant gravity theory~\cite{Gao:2014soa,Gao:2014fra,Gao:2018znj}.
The spatially covariant gravity theory includes the ghost condensation~\cite{Arkani-Hamed:2003pdi}, Ho\v{r}ava-Lifsitz gravity~\cite{Horava:2008ih,Horava:2009uw}, and its infrared limit, the khronometric theory~\cite{Blas:2009yd}, as well as the effective field theory (EFT) of inflation/dark energy~\cite{Cheung:2007st,Gubitosi:2012hu} as concrete examples.
Moreover, there exists an interesting family of the spatially covariant gravity theory having only two tensorial degrees of freedom (TTDOF) and no scalar propagating mode, which has drawn much attention recently~\cite{Lin:2017oow,Iyonaga:2018vnu,Gao:2019twq,Lin:2020nro,Iyonaga:2021yfv}.

To test modified gravity, it is important to investigate the behavior of gravity in the Newtonian and post-Newtonian (PN) regimes, as well as the complementary cosmological regime.
In the Horndeski family of scalar-tensor theories, the Vainshtein mechanism~\cite{Vainshtein:1972sx} plays a crucial role in screening the extra force mediated by the scalar DOF with the help of its nonlinear derivative interactions and thus recovering GR in the quasi-static regime on small scales~\cite{Kimura:2011dc} (see~\cite{Babichev:2013usa} for a review of the Vainshtein mechanism).
In DHOST theories, the partial breaking of the Vainshtein screening occurs in a material body, providing possible tests of such theories~\cite{Kobayashi:2014ida,Crisostomi:2017lbg,Langlois:2017dyl,Dima:2017pwp,Hirano:2019scf,Crisostomi:2019yfo}.
In the U-DHOST family of theories, the Newtonian and PN regimes are more subtle.
As reported in~\cite{Kobayashi:2023lyt}, Lorentz breaking terms can be dominant as compared to nonlinear derivative interactions in generic U-DHOST theories, leading to the potential recovery of GR by the choice of parameters rather than with the Vainshtein mechanism.

The parametrized post-Newtonian (PPN) formalism~\cite{Will:2005va,Will:2018bme} is a convenient tool to see whether or not a given metric theory of gravity is consistent with experiments in the PN limit, and has been widely used in the literature.
In the PPN formalism, the metric in the PN limit is characterized by several parameters, and their deviations from the values in GR are bounded by experiments.
In~\cite{Kobayashi:2023lyt}, only one of the PPN parameters (the one usually denoted as $\gamma$) in generic U-DHOST theories has been determined and the potential recovery of GR has been speculated.
In this paper, we determine all the PPN parameters in U-DHOST theories, quantify their deviations from GR in the PN limit, and provide experimental tests of them.

Cosmological probes offer us complementary tests of modified gravity on large scales~\cite{Jain:2010ka,Joyce:2014kja,Koyama:2015vza,Arai:2022ilw}.
In this context, the infinite functional freedom in Horndeski and beyond Horndeski theories can be conveniently incorporated into the EFT framework for dark energy/modified gravity with a finite number of EFT parameters~\cite{Gubitosi:2012hu,Gleyzes:2013ooa,Bellini:2014fua,Langlois:2017mxy}.
Each EFT parameter has a physically transparent interpretation and can be constrained by current and future cosmological observations~\cite{Arai:2022ilw}.
The main goal of this paper is to connect the two useful parametrizations of modified gravity, i.e. the PPN and EFT parameters, within the framework of U-DHOST theories,
which allows us to translate experimental bounds on the PPN parameters to constraints on U-DHOST theories in terms of the EFT of dark energy parameters.
Our study shares a similar spirit with~\cite{Lombriser:2018guo,Renevey:2020tvr}, but our interest is in scalar-tensor theories that are not equipped with the Vainshtein screening mechanism.

This paper is organized as follows.
In the next section, we review the U-DHOST family of scalar-tensor theories and introduce the EFT parameters as an economic description of scalar-tensor theories having infinite functional freedom.
In Sec.~\ref{PPN_formalism}, we start with a brief review of the PPN formalism and then present the form of the metric and scalar field expanded to the required PN order.
In Sec.~\ref{PPN_parameters:U-DHOST}, all the PPN parameters in U-DHOST theories are determined.
In Sec.~\ref{sec:examples}, we apply our general results to two concrete examples of scalar-tensor theories.
In Sec.~\ref{sec:constraints_EFT para}, we derive constraints on the EFT parameters from experimental bounds on the PPN parameters.
We also present a U-DHOST theory whose PPN parameters are the same as those in GR.
Finally, we draw our conclusions in Sec.~\ref{conclusions}.

\section{U-DHOST theory and EFT of dark energy parameters}\label{U-DHOST}

In this section,
we consider the U-DHOST family of scalar-tensor theories~\cite{DeFelice:2018ewo}, which is a class of higher-order scalar-tensor theories satisfying the degeneracy conditions when restricted to the unitary gauge.
Even though the theory is not fully degenerate, it is free from ghosts
because, away from the unitary gauge, the potentially dangerous additional mode obeys an elliptic equation rather than a hyperbolic equation
and hence is not a propagating DOF.
The non-propagating DOF also refers to a mode with an infinite speed of sound.
Such a mode is called a generalized instantaneous mode or a shadowy mode (see~\cite{DeFelice:2021hps} for more detailed discussions).

The action we consider is given by
\begin{align}
    S &=S_{\text{grav}}+S_{\text{m}},
\end{align}
where $S_{\text{grav}}$ is the action for higher-order scalar-tensor theories,
\begin{align}
            S_{\text{grav}} 
            =\int\D^4x\sqrt{-g}\biggl[P+
                f\mathcal{R}+
                A_1\phi_{\mu\nu}\phi^{\mu\nu}
                +A_2(\Box\phi)^2 
                +A_3\Box\phi \phi^\mu\phi^\nu\phi_{\mu\nu}
                +A_4\phi^\mu\phi_{\mu\rho}\phi^{\rho\nu}\phi_\nu 
                +A_5(\phi^\mu\phi^\nu\phi_{\mu\nu})^2
        \biggr],\label{eq:U-DHOST-action}
\end{align}
with notations $\phi_\mu:=\nabla_\mu\phi$ and $\phi_{\mu\nu}:=\nabla_\mu\nabla_\nu\phi$,
and $S_{\text{m}}$ is the action for matter fields
which are assumed to be minimally coupled to gravity.
In the gravitational part of the action, $f$, $P$, and $A_I$ ($I=1,2,3,4,5$) are functions of the scalar field $\phi$ and its canonical kinetic term, $X:=-g^{\mu\nu}\phi_\mu\phi_\nu/2$.
One may also add to the Lagrangian the term of the form $Q(\phi,X)\Box\phi$,
which we do not consider for simplicity in this paper.

Since the action~\eqref{eq:U-DHOST-action} gives rise to higher-derivative field equations in general, we must impose some restrictions on the functions $f$ and $A_I$ so that the system is degenerate and hence is free from ghosts.
(The $k$-essence term $P$ has nothing to do with the degeneracy of the system.)
In Ref.~\cite{Langlois:2015cwa}, the full set of degeneracy conditions for the action~\eqref{eq:U-DHOST-action} was derived with the detailed classification of fully degenerate theories.
Later it was noticed that one can still have ghost-free theories even if the degeneracy conditions are satisfied only when restricted to the unitary gauge~\cite{DeFelice:2018ewo}.
In the latter version of healthy higher-derivative scalar-tensor theories,
five of the six functions ($f$ and $A_I$) are independent and
$A_1,\cdots, A_5$ are given in the form~\cite{DeFelice:2018ewo}
\begin{align}
     A_1&=a_1-\frac{f}{2X},\quad A_2=a_2+\frac{f}{2X},\quad A_3=\frac{f}{2X^2}-\frac{f'}{X}+2a_1a_3+2\left(3a_3+\frac{1}{2X}\right)a_2,
     \notag \\
       A_4&=a_4+\frac{f'}{2X}-\frac{f}{2X^2}+\frac{a_1}{X},\quad A_5=\frac{a_4}{2X}-\frac{f'}{4X^2}+a_1\left(\frac{1}{4X^2}+3a_3^2+\frac{a_3}{X}\right)+a_2\left(3a_3+\frac{1}{2X}\right)^2, \label{eq:U-degenerate-functions}
\end{align}
where $f$ and $a_I$ ($I=1,2,3,4$) are arbitrary functions of $\phi$ and $X$.
The action for a U-DHOST theory is thus given by Eq.~\eqref{eq:U-DHOST-action} with the functions~\eqref{eq:U-degenerate-functions}.
The U-DHOST family includes khronometric~\cite{Blas:2009yd} and TTDOF~\cite{Gao:2019twq} theories as specific examples, as well as all scalar-tensor theories in the Horndeski/DHOST family.
(See Sec.~\ref{sec:examples} for the concrete form of the functions corresponding to the khronometric and TTDOF theories.)

To classify and characterize (U-)DHOST/Horndeski theories, it is convenient to introduce the following EFT of dark energy parameters~\cite{Bellini:2014fua,Langlois:2017mxy}:
\begin{align}
           &M^2=2(f+2XA_1),
        \quad 
        M^2(1+\alpha_T)=2f,
        \quad 
        M^2(1+\alpha_H)=2(f-2Xf'),\notag 
         \\ &
        M^2\left(1+\frac{2}{3}\alpha_L\right)=2(f-2XA_2),
        \quad 
        M^2\beta_1=2\left[Xf'-X(A_2-XA_3)\right],\notag 
         \\ & 
        M^2\beta_2=4X\left[
                A_1+A_2-2X(A_3+A_4)+4X^2A_5 
        \right],\quad
        M^2\beta_3=-8\left[Xf'+X(A_1-XA_4)\right],
\end{align}
where a prime denotes differentiation with respect to $X$.
In the U-DHOST family, we have the relation among the EFT parameters which follows from the degeneracy condition,
\begin{align}
    \beta_2=-\frac{6\beta_1^2}{1+\alpha_L},\label{eq:u-deg-condition}
\end{align}
but aside from this the EFT parameters are independent.
Each EFT parameter has some physical significance.
For example, $\alpha_T$ characterizes the deviation of the propagation speed of gravitational waves $\cgw$ from that of light,
\begin{align}
    \alpha_T=\cgw^2-1,
\end{align}
and $M^{-2}$ is related to the effective gravitational coupling for gravitational waves as\footnote{Gravitational waves $h_{ij}$ obey $\cgw^{-2}\partial_t^2h_{ij}-\nabla^2h_{ij} \sim 16\pi \ggw T_{ij}$ in the presence of the energy-momentum tensor.}
\begin{align}
    M^{-2}=8\pi \ggw \cgw^2.\label{def:ggw}
\end{align}
In a cosmological setup, the field $\phi$ varies on cosmological time scales, and so do those EFT parameters.
Numerical codes are available for linear cosmology of modified gravity in terms of the EFT parameters~\cite{Zumalacarregui:2016pph,Hiramatsu:2020fcd}, and
observational constraints on scalar-tensor theories are often translated to the limits on the EFT parameters.

For later purposes,
we introduce two additional parameters as
\begin{align}
    M^2\delta_1&:=2X(f-2Xf')',\label{def:delta1}
    \\ 
    M^2\delta_2&:=-8X\left[X(f'+A_1-XA_4)\right]'.\label{def:delta2}
\end{align}
While these parameters do not appear in the analysis of linear cosmology,
we will encounter them when working in the PN approximation.

Throughout the paper we assume that
\begin{align}
    \alpha_L\neq 0,
\end{align}
which is typically the case in Lorentz-breaking theories such as the khronometric theory~\cite{Blas:2009yd}.
This is crucial for the validity of our calculations.
It was shown in Ref.~\cite{Kobayashi:2023lyt} that, if $\alpha_L\neq 0$, we do not need to take into account the Vainshtein mechanism arising from nonlinear derivative interactions in the weak-field regime.
Thanks to this, the PN calculations in the present paper are feasible.

Before closing this section, it should be emphasized that we will not impose the degeneracy conditions among the functions in the Lagrangian in almost all the calculations in the present paper, even though we are interested in ghost-free theories satisfying those conditions in the unitary gauge.
In particular, we do not need to impose the relation~\eqref{eq:u-deg-condition} to push forward our PN calculations.
The reason is that the unitary degeneracy condition is used to remove dangerous higher \textit{time} derivatives in the field equations, but in the PN approximation, such terms are assumed to be small from the beginning and hence are dropped from the equations even without imposing the unitary degeneracy condition.

\section{The PPN formalism and the PN expansion in U-DHOST theory}\label{PPN_formalism}

In this section, we introduce the PPN formalism,
which describes deviations from GR in the PN limit with a finite number of potentials and parameters.
(See the textbook~\cite{Will:2018bme} for an extensive review of the PPN formalism.)
After that, we present the PN expansion of the metric and the scalar field in U-DHOST theories, assuming that the gradient of the scalar field is timelike.
This assumption on the profile of the scalar field makes a crucial difference between the PN expansion in old-fashioned scalar-tensor theories (such as the Brans-Dicke theory)~\cite{Will:2018bme} and the present case.

\subsection{The energy-momentum tensor}

Let us first introduce the energy-momentum tensor of matter.
We assume that matter is given by a perfect fluid
with the energy-momentum tensor
\begin{align}
    T^{\mu\nu}=(\rho+\rho\Pi+p)u^\mu u^\nu+p g^{\mu\nu},\label{emtensor:fluid}
\end{align}
where $\rho$ is the rest-mass density, $\rho \Pi$ is the density of internal kinetic and thermal energy, $p$ is the isotropic pressure, and $u^\mu=u^0(1,v^i)$ is the four-velocity of the fluid element.
The energy-momentum tensor satisfies the conservation equation
\begin{align}
    \nabla_\mu T^{\mu\nu}=0.
\end{align}
In addition, we assume the baryon number conservation,
\begin{align}
    \nabla_\mu (\rho u^\mu)
    =0 
    \quad \Rightarrow \quad 
    \partial_t \rho^*+\partial_i\left(\rho^* v^i\right)=0,\label{eq:conserv}
\end{align}
where
\begin{align}
    \rho^*:=\sqrt{-g}\rho u^0
\end{align}
is an auxiliary density variable called the conserved density.

\subsection{The post-Newtonian bookkeeping}

Next, we introduce a bookkeeping parameter $\epsilon \,(\ll 1)$
to track the smallness of various quantities in the PN approximation.
We assume that the velocity of an element of the fluid is a small quantity
of order $\epsilon$:
\begin{align}
    v=\mathcal{O}(\epsilon).
\end{align}
The Newtonian gravitational potential is defined as
\begin{align}
    U(t,\vec{x}):=\gn \int\mathrm{d}^3 y \frac{\rho^*(t,\vec{y})}{|\vec{x}-\vec{y}|}
    \quad \Leftrightarrow\quad \Delta U=-4\pi \gn \rho^*,
    \label{def:newtonian-potential}
\end{align}
where $\gn$ is the Newtonian gravitational constant and $\Delta$ is the flat-space Laplacian.
From virial relations we have
\begin{align}
    U\sim v^2=\mathcal{O}(\epsilon^2).
\end{align}
Then, assuming that the gravitational force is balanced by the pressure gradient, i.e. $\rho \partial_i U = \partial_i p$, we get
\begin{align}
    p/\rho\sim U= \mathcal{O}(\epsilon^2).
\end{align}
Thermodynamics tells us that
the internal energy $E$ of a system of volume $\mathcal{V}$ is related to the pressure as $p\mathcal{V} \sim E \sim \rho \Pi \mathcal{V}$,
and hence
\begin{align}
    \Pi \sim p/\rho = \mathcal{O}(\epsilon^2).
\end{align}
The PN order of the potential and fluid variables will be increased by taking a time derivative as follows
\begin{align}
    \partial_t \sim v^i\partial_i = \mathcal{O}(\epsilon\partial_i).
\end{align}
Note, however, that the time derivative(s) of the EFT parameters will be ignored in the analysis of this paper because they are assumed to vary on cosmological time scales and hence their time derivative yields the order of the Hubble parameter which is much smaller than $\mathcal{O}(\epsilon\partial_i)$.
In other words, we will work in the action~\eqref{eq:U-DHOST-action} with the shift symmetry whose functions are only dependent on $X$.

\subsection{Post-Newtonian expansion of the metric and the scalar field}

By adopting the standard PPN gauge, the PN expansions of the metric and scalar fields are given as follows:
\begin{align}
    g_{00}&=-1+2U+2 \left(\psi-\ppnb U^2\right)+\mathcal{O}(\epsilon^6),\label{g_00}\\
    g_{0i}&=B_i+\mathcal{O}(\epsilon^5),\\
    g_{ij}&=\left(1+2\ppng U+2 C\right)\delta_{ij}+D_{ij}+\mathcal{O}(\epsilon^6),\label{g_ij}
\end{align}
where 
\begin{align}
        \psi&:=\frac{1}{2}\left(2\ppng+1+\ppna{3}+\ppnz{1}-2\ppnxi\right)\Phi_1+\left(1-\ppnb+\ppnz{2}+\ppnxi\right)\Phi_2
        \notag \\ 
        &\quad+\left(1+\ppnz{3}\right)\Phi_3+
        \left(3\ppng+3\ppnz{4}-2\ppnxi\right)\Phi_4-\frac{1}{2}
        \left(\ppnz{1}-2\ppnxi\right)\Phi_6-\ppnxi \Phi_W
        \notag \\ &=\mathcal{O}(\epsilon^4)\label{psi},\\
        B_i&:=-\left[2\left(1+\ppng\right)+\frac{\ppna{1}}{2}\right]V_i-\frac{1}{2}\left(1+\ppna{2}-\ppnz{1}+2\ppnxi\right)\partial_i \dot{\mathcal{X}}
        \notag \\ & =\mathcal{O}(\epsilon^3)\label{B_i},
        \\
        C&:=d_{UU}U^2+d_W \Phi_W+d_1 \Phi_1+d_2 \Phi_2+d_3 \Phi_3+d_4 \Phi_4+d_6 \Phi_6+d_{\mathcal{X}}\ddot{\mathcal{X}}
        \notag \\ &=\mathcal{O}(\epsilon^4)\label{C},
\end{align}
and $D_{ij}$ is a quantity of $\mathcal{O}(\epsilon^4)$ satisfying
the traceless condition, i.e. $\delta^{ij}D_{ij}=0$. 
Here and hereafter, a dot denotes the derivative with respect to $t$, and the potentials in Eqs.~\eqref{psi}--\eqref{C} are defined by
\begin{align}
    \mathcal{X}&=\gn \int \mathrm{d}^3y \rho^*(t,\vec{y}) |\vec{x}-\vec{y}|,\label{chi}\\
    V_i&=\gn \int \mathrm{d}^3y \frac{\rho ^*(t,\vec{y}) v_i(t,\vec{y})}{|\vec{x}-\vec{y}|},\\
    \Phi_1&=\gn \int \mathrm{d}^3y \frac{\rho^*(t,\vec{y}) v^2(t,\vec{y})}{|\vec{x}-\vec{y}|},\\
    \Phi_2&=\gn \int \mathrm{d}^3y \frac{\rho^*(t,\vec{y}) U(t,\vec{y})}{|\vec{x}-\vec{y}|},\\
    \Phi_3&=\gn \int \mathrm{d}^3y \frac{\rho^*(t,\vec{y}) \Pi(t,\vec{y})}{|\vec{x}-\vec{y}|},\\
    \Phi_4&=\gn \int \mathrm{d}^3y \frac{p(t,\vec{y})}{|\vec{x}-\vec{y}|},\\
    \Phi_6&=\gn \int \mathrm{d}^3y \frac{\rho^*(t,\vec{y})\left[\vec{v}(t,\vec{y})\cdot(\vec{x}-\vec{y})\right]^2}{|\vec{x}-\vec{y}|^3},\\
    \Phi_W&=\gn \int \mathrm{d}^3y\mathrm{d}^3z\rho^*(t,\vec{y})\rho^*(t,\vec{z})\frac{(\vec{x}-\vec{y})}{|\vec{x}-\vec{y}|^3}\cdot\left[\frac{(\vec{y}-\vec{z})}{|\vec{x}-\vec{z}|}-\frac{(\vec{x}-\vec{z})}{|\vec{y}-\vec{z}|}\right]\label{Phi_W}.
\end{align}
The constants $\ppng,\ppnb,\ppna{1},\ppna{2},\ppna{3},\ppnz{1},\ppnz{2},\ppnz{3},\ppnz{4}$, and $\ppnxi$ are referred to as the PPN parameters.
One could include $\ddot{\mathcal{X}} \,[=\mathcal{O}(\epsilon^4)]$ in $g_{00}$ and $\partial_i\partial_j\mathcal{X}\,[=\mathcal{O}(\epsilon^2)]$ in $g_{ij}$.
However, they have been removed by using an infinitesimal coordinate transformation of the form
\begin{align}
    t\to t+\lambda_1\dot{\mathcal{X}},
    \quad x^i\to x^i+\lambda_2\partial^i\mathcal{X},
    \label{cood-tr-unit-to-PPN}
\end{align}
where $\lambda_{1}$ and $\lambda_{2}$ are two constants.
This gauge choice is what we have been calling the standard PPN gauge.

Note that the potentials in Eqs.~\eqref{chi}--\eqref{Phi_W} are not all independent.
In fact, from the conservation of the baryon number~\eqref{eq:conserv}, one can find
\begin{align}
    \dot U+\partial_iV_i=0,\label{relation-01}
\end{align}
and, by definition, we also have
\begin{align}
    \Delta \mathcal{X}&=2U,\label{relation-02}
    \\ 
    \Delta\Phi_2&=U\Delta U.\label{relation-03}
\end{align}
We will use these relations in our calculations.
In the 1PN approximation, $g_{ij}$ is expanded up to $\mathcal{O}(\epsilon^2)$, and hence usually we do not need to introduce the $\mathcal{O}(\epsilon^4)$ terms in $g_{ij}$, i.e. $C$ and $D_{ij}$.
In the present case, however, the terms in $C$ with the coefficients $d_{UU},\dots,d_{\mathcal{X}}$ will be used at an intermediate step in determining the PPN parameters.
This is due to the scalar field configuration with timelike gradient.
On the other hand, the traceless part $D_{ij}$ is completely irrelevant to our analysis.

Far from the post-Newtonian source, the scalar field is assumed to be of the form $\phi=q t$ and thus $X=q^2/2$, where $q$ is a constant.
When the gradient of the scalar field is timelike,
a natural and convenient gauge choice would be the unitary gauge, in which the scalar field remains $\phi=qt$ everywhere to the required PN order. 
However, in the above we have already chosen to work in the standard PPN gauge,
and therefore $\phi$ must also be expanded in powers of $\epsilon$.\footnote{This is only a matter of the gauge choice.
One can instead work in the unitary gauge, but then there must be an extra term proportional to $\ddot{\mathcal{X}}$ in $g_{00}$. The PN expansions in the unitary gauge have been addressed in similar modified gravity theories in~\cite{Blas:2011zd,Lin:2013tua,Qiao:2021fwi}.}
The coordinate transformation~\eqref{cood-tr-unit-to-PPN} used to move to the standard PPN gauge implies that the scalar field in that gauge is written as
\begin{align}
\phi=q\left(t+\ppngx\dot{\mathcal{X}}\right)+\mathcal{O}(\epsilon^5),
\label{PN-ex-phi}
\end{align}
where $\ppngx$ is a dimensionless constant.
Thus, the leading correction is assumed to be of $\mathcal{O}(\epsilon^3)$.
The configuration of the scalar field~\eqref{PN-ex-phi} should be contrasted with that in the previous literature where the gradient of the scalar field is spacelike (see, e.g.,~\cite{Will:2018bme}).
As we will see in the next section, Eq.~\eqref{PN-ex-phi} indeed gives a consistent ansatz in the present case.
See also Appendix~\ref{app:aL} for a further discussion on this point.

A remark is in order.
To derive correctly a weak gravitational field around a Newtonian source in the Horndeski/DHOST family of scalar-tensor theories, it is necessary to take into account nonlinear derivative interactions of metric and scalar-field fluctuations which are responsible for the Vainshtein screening mechanism (and its partial breaking)~\cite{Kimura:2011dc,Kobayashi:2014ida,Crisostomi:2017lbg,Langlois:2017dyl,Dima:2017pwp,Hirano:2019scf,Crisostomi:2019yfo}.
In general, we expect that the metric and the scalar field are expanded differently and the PPN parameters may depend on the distance from the source in scalar-tensor theories equipped with the Vainshtein screening mechanism.
However, in scalar-tensor theories with $\alpha_L\neq 0$, we do not need to care about nonlinear derivative interactions and thus can avoid the complexity introduced by them~\cite{Kobayashi:2023lyt}.
For this reason, we may assume that the metric is expanded in the usual way as in the conventional PPN formalism. See~\cite{Lombriser:2018guo} for the PPN parameters in screened Horndeski gravity.

\section{PPN parameters in U-DHOST theory}\label{PPN_parameters:U-DHOST}

We now determine the PPN parameters in the U-DHOST family of theories by substituting the metric and the scalar field presented in the previous section into the field equations and collecting the terms at each order of $\epsilon$.
In doing so, we do not need to impose the unitary degeneracy condition~\eqref{eq:u-deg-condition} and hence we treat $\beta_2$ as if it were a free parameter.
However, we do assume that $\alpha_L\neq 0$.

The field equations for the metric and the scalar field are given respectively by
\begin{align}
    \frac{2}{\sqrt{-g}}\frac{\delta S}{\delta g^{\mu\nu}}=\mathcal{E}_{\mu\nu}-T_{\mu\nu}=0,
\end{align}
and
\begin{align}
    \frac{1}{\sqrt{-g}}\frac{\delta S}{\delta \phi}=\mathcal{E}_{\phi}=0.
\end{align}
Since we have the identity $\nabla^\nu\mathcal{E}_{\mu\nu}=-\phi_\mu\mathcal{E}_\phi$ which follows from the general covariance of the theory, we will only use the gravitational field equations $\mathcal{E}_{\mu\nu}=T_{\mu\nu}$ in the following analysis. The scalar field equation $\mathcal{E}_\phi=0$ is then automatically satisfied.
The explicit expression for $\mathcal{E}_{\mu\nu}$
is presented in Appendix~\ref{app:A}.

A direct calculation shows that $\mathcal{E}_{\mu\nu}$ is expanded as
\begin{align}
    \mathcal{E}_{00}&=\mathcal{E}_{00}^{(0)}+\mathcal{E}_{00}^{(2)}\epsilon^2+
    \mathcal{E}_{00}^{(4)}\epsilon^4+\mathcal{O}(\epsilon^6),
    \\ 
    \mathcal{E}_{0i}&=\mathcal{E}_{0i}^{(3)}\epsilon^3+\mathcal{O}(\epsilon^5),
    \\ 
    \mathcal{E}_{ij}&=\mathcal{E}_{ij}^{(0)}+\mathcal{E}_{ij}^{(2)}\epsilon^2+
    \mathcal{E}_{ij}^{(4)}\epsilon^4+\mathcal{O}(\epsilon^6).
\end{align}
The energy-momentum tensor~\eqref{emtensor:fluid} can be written explicitly in terms of the potentials as
\begin{align}
    T_{00}&=T_{00}^{(2)}\epsilon^2+T_{00}^{(4)}\epsilon^4+\mathcal{O}(\epsilon^6),
    \\ 
    T_{0i}&=T_{0i}^{(3)}\epsilon^3+\mathcal{O}(\epsilon^5),
    \\ 
    T_{ij}&=T_{ij}^{(4)}\epsilon^4+\mathcal{O}(\epsilon^6),
\end{align}
where
\begin{align}
    T_{00}^{(2)}&=\rho^*=-\frac{\Delta U}{4\pi \gn},
    \\ 
    T_{00}^{(4)}&=\rho^*\left(
    \frac{v^2}{2}-2U-3\ppng U+\Pi
    \right)=-\frac{1}{4\pi\gn}\left[\frac{\Delta\Phi_1}{2}-\left(2+3\ppng\right)\Delta\Phi_2
    +\Delta\Phi_3\right],
    \\ 
    T_{0i}^{(3)}&=-\rho^*\delta_{ij}v^j=\frac{\Delta V_i}{4\pi\gn}.
\end{align}
The $\mathcal{O}(\epsilon^4)$ part of the $(ij)$ component of the energy-momentum tensor is given by $T_{ij}^{(4)}=\rho^*v_iv_j+p \delta_{ij}$, but we only use its trace:
\begin{align}
    \delta^{ij}T_{ij}^{(4)}=-\frac{1}{4\pi\gn}\left(\Delta\Phi_1+3\Delta\Phi_4\right).
\end{align}

We now proceed to determine the PPN parameters using the field equations at each order in $\epsilon$.
The readers who are not interested in technical details can skip the derivation and directly refer to Table~\ref{table:PPN-paras} for the summary of the main results.

\subsection{$\mathcal{O}(\epsilon^0)$ equations}\label{subsec:0}

The zeroth-order equations are used to set our cosmological boundary conditions.
We have 
\begin{align}
    \mathcal{E}_{00}^{(0)}&=P-q^2P'=0,\label{0th:00}
    \\ 
    \mathcal{E}_{ij}^{(0)}&=-P\delta_{ij}=0.\label{0th:ij}
\end{align}
Here and hereafter, the functions of $X$ are evaluated at $X=q^2/2$.
We assume that the function $P$ is such that Eqs.~\eqref{0th:00} and~\eqref{0th:ij} are satisfied for some $q$.
The value of $q$ is thus determined.

\subsection{$\mathcal{O}(\epsilon^2)$ equations}\label{subsec:2}

At $\mathcal{O}(\epsilon^2)$, we obtain the expression for $\gn$ and $\ppng$.
A straightforward calculation yields
\begin{align}
    \mathcal{E}_{00}^{(2)}&=
    -q^4P''U+
    2\left\{-2\ppng f+q^2\left[-2A_1+q^2A_4-2\left(1-\ppng\right)f'
    \right]\right\}\Delta U
    =T_{00}^{(2)},\label{Eq:2nd}
    \\ 
    \mathcal{E}_{ij}^{(2)}&=2\left[(1-\ppng)f-q^2f'\right]
    \left(\partial_i\partial_j-\delta_{ij}\Delta\right)U=0.
    \label{TrEq:2nd}
\end{align}
The first term in Eq.~\eqref{Eq:2nd} plays the role of a mass term for $U$.
We simply assume that $P$ is such that this term can be ignored relative to the other terms.
Otherwise, $U$ would not be given by the Newtonian gravitational potential~\eqref{def:newtonian-potential}.

Using Eq.~\eqref{TrEq:2nd} one obtains
\begin{align}
    \ppng&=1-\frac{q^2f'}{f}=\frac{1+\alpha_H}{1+\alpha_T},\label{gamma}
\end{align}
and then from Eq.~\eqref{Eq:2nd} one gets
\begin{align}
    8\pi \gn&=\frac{f}{2(f-q^2f')^2+q^2f(2f'+2A_1-q^2A_4)}=\left[\frac{2(1+\alpha_T)}{2(1+\alpha_H)^2-(1+\alpha_T)\beta_3}\right]\frac{1}{M^2}.\label{G_Newton}
\end{align}
This result reproduces that of the linear perturbation analysis of~\cite{Kobayashi:2023lyt}.
In terms of the effective gravitational coupling for gravitational waves defined by Eq.~\eqref{def:ggw},
we have
\begin{align}
    \gn=\left[\frac{2(1+\alpha_T)^2}{2(1+\alpha_H)^2-(1+\alpha_T)\beta_3}\right]
    \ggw.
\end{align}

\subsection{$\mathcal{O}(\epsilon^3)$ equations}\label{subsec:3}

Next, we investigate the $\mathcal{O}(\epsilon^3)$ equation.
The $(0i)$ component of the field equations yields the equation at this order:
\begin{align}
    \mathcal{E}_{0i}^{(3)}
    &=\frac{1}{2}\left(\mu_{V}
    \Delta V_i+\tilde\mu_V\partial_i\partial_j V_j\right)=T_{0i}^{(3)},
\end{align}
with 
\begin{align}
    \mu_V&:=\left[4\left(1+\ppng\right)+\ppna{1}\right]\left(f+q^2A_1\right)
    \notag \\ & \quad =\frac{M^2}{2}\left[4\left(1+\ppng\right)+\ppna{1}\right],
    \\ 
    \tilde\mu_V&:=-\left(4-4\ppng+\ppna{1}\right)f+4q^2f'
    +q^2\left(8\ppngx+\ppna{1}-4\ppna{2}+4\ppnz{1}-8\ppnxi\right)A_1
    \notag \\
    &\quad
    +2q^2\left(-2\ppng+4\ppngx +\ppna{1}-2\ppna{2}+2\ppnz{1}-4\ppnxi\right)
    A_2+2q^4A_3
    \notag \\ & \quad =M^2\left[-2\left(1-\ppng\right)-\frac{\ppna{1}}{2}+4\beta_1
    -\frac{2\alpha_L}{3}\left(
    2-2\ppng+\ppna{1}-2\ppna{2}+4\ppngx+2\ppnz{1}-4\ppnxi
    \right)\right].
\end{align}
Here, we used Eqs.~\eqref{relation-01} and~\eqref{relation-02} to express $\mathcal{E}_{0i}^{(3)}$ solely in terms of $V_i$.
By comparing the coefficients of $\Delta V_i$ and $\partial_i\partial_jV_j$ on the left-hand side with those on the right-hand side, we arrive at the equations
\begin{align}
    \mu_V=\frac{1}{2\pi\gn},\quad \tilde\mu_V=0.
\end{align}
Using the former equation, we can determine $\ppna{1}$ as
\begin{align}
    \ppna{1}
    =4\left[\frac{2(1+\alpha_H)^2}{1+\alpha_T}-\frac{1+\alpha_H}{1+\alpha_T}-1-\beta_3\right].
\end{align}
It is convenient to express $\ppna{1}$ using $\ppng$ and $\cgw$ as 
\begin{align}
    \ppna{1}=4\left[2\left(\ppng\right)^2\cgw^2-\ppng-1-\beta_3\right].
\end{align}
From $\tilde\mu_V=0$ we obtain 
\begin{align}
    \ppna{2}-2\ppngx-\ppnz{1}+2\ppnxi
    =\frac{4(1+\alpha_H)^2}{1+\alpha_T}-\frac{3(1+\alpha_H)}{1+\alpha_T}-1-2\beta_3+\frac{3}{\alpha_L}\left[\frac{\alpha_H(1+\alpha_H)}{1+\alpha_T}-\beta_1-\frac{\beta_3}{2}\right].\label{soln:a2}
\end{align}
At this step one can only express the combination
$\ppna{2}-2\ppngx-\ppnz{1}+2\ppnxi$ in terms of the EFT parameters.
To determine each of the four parameters,
we need to proceed and use the $\mathcal{O}(\epsilon^4)$ equations. 

\subsection{$\mathcal{O}(\epsilon^4)$ equations}\label{subsec:4}

Let us finally discuss the $\mathcal{O}(\epsilon^4)$ equations.
The rest of the PPN parameters can be determined at this order.
The basic procedure is tedious but straightforward.
We first write $\mathcal{E}_{00}^{(4)}$ and $\delta^{ij}\mathcal{E}_{ij}^{(4)}$ explicitly in terms of the eight potentials $\Delta\Phi_{1,2,3,4,6,W}$, $\Delta U^2$, and $\Delta\ddot{\mathcal{X}}$.
In doing so, we rewrite $\partial_iU\partial_iU$, $U\Delta U$, $\ddot U$, and $\partial_i\dot V_i$ appearing at intermediate steps in terms of $\Delta\Phi_2$, $\Delta U^2$, and $\Delta\ddot{\mathcal{X}}$ by using Eqs.~\eqref{relation-01}--\eqref{relation-03} and the identity
$\partial_iU\partial_iU=\Delta U^2/2-U\Delta U$.
We then compare the coefficients of the eight potentials on the left-hand sides of $\mathcal{E}_{00}^{(4)}=T_{00}^{(4)}$ and $\delta^{ij}\mathcal{E}_{ij}^{(4)}=\delta^{ij}T_{ij}^{(4)}$ with those on their right-hand sides.
We now have $8\times 2 = 16$ independent equations at $\mathcal{O}(\epsilon^4)$ for the 17 parameters yet to be determined ($d_{UU,W,1,2,3,4,6,\mathcal{X}},\ppnb,\ppna{2,3},\ppnz{1,2,3,4},\ppnxi$, and $\ppngx$).
With the help of Eq.~\eqref{soln:a2} obtained at $\mathcal{O}(\epsilon^3)$,
we can determine all these 17 parameters in the end.
We do not need to care about the traceless part of the $(ij)$ component of the field equations at $\mathcal{O}(\epsilon^4)$ because it is used to determine $D_{ij}$ which is irrelevant to the PN limit.

Explicitly, we have
\begin{align}
    M^{-2}\mathcal{E}_{00}^{(4)}&= \mu_C\Delta C+
    \mu_1\Delta\Phi_1 + \mu_2\Delta\Phi_2
    +\mu_3\Delta\Phi_3
    +\mu_4 \Delta\Phi_4 
    +\mu_6\Delta\Phi_6
    +\mu_W\Delta\Phi_W+\mu_{UU}\Delta U^2 
    +\mu_{\mathcal{X}}\Delta\ddot{\mathcal{X}},
    \\
    M^{-2}\delta^{ij}\mathcal{E}_{ij}^{(4)}&=
    \nu_C\Delta C+
    \nu_1\Delta\Phi_1 + \nu_2\Delta\Phi_2
    +\nu_3\Delta\Phi_3
    +\nu_4 \Delta\Phi_4 
    +\nu_6\Delta\Phi_6
    +\nu_W\Delta\Phi_W+\mu_{UU}\Delta U^2 
    +\nu_{\mathcal{X}}\Delta\ddot{\mathcal{X}},
\end{align}
where
\begin{align}
    &\mu_{C}:=-2\cgw^2\ppng,
    \quad 
    \mu_1:=\frac{\beta_3}{2}\left(1+\ppna{3}+2\ppng+\ppnz{1}-2\ppnxi\right),
    \notag 
    \\ & 
    \mu_2:=\cgw^2(\ppng)^2\left(4+5\ppng\right)-4\ppng\delta_1+\delta_2 
    +\beta_3\left(-\frac{1}{2}-2\ppnb-3\ppng+\ppnz{2}+\xi \right),
    \notag 
    \\ & 
    \mu_3:=\beta_3\left(1+\ppnz{3}\right),
    \quad 
    \mu_4:=\beta_3\left[3\left(\ppng+\ppnz{4}\right)-2\ppnxi\right],
    \quad 
    \mu_6:=\beta_3\left(-\frac{\ppnz{1}}{2}+\ppnxi\right),
    \notag
    \\ & 
    \mu_W:=-\beta_3\ppnxi,\quad 
    \mu_{UU}:=\frac{1}{4}
    \left[6\cgw^2(\ppng)^3+\beta_3\left(3-4\ppnb+2\ppng\right)+2\delta_2\right],
    \notag 
    \\ & 
    \mu_{\mathcal{X}}:=
    \frac{1}{2}\left(\beta_2+2\ppngx \beta_3\right)
    +\frac{3\beta_1}{2\alpha_L}\left[2\beta_1+\beta_3+2\ppng\left(1-\cgw\ppng\right)\right],
    \notag
    \\ & 
    \nu_C:=2\cgw^2,\quad 
    \nu_1:=-\cgw\ppng\left(1+\ppna{3}+2\ppng+\ppnz{1}-2\ppnxi\right),
    \notag
    \\ & 
    \nu_2:=\frac{1}{2}\left\{
        \beta_3-2\cgw\ppng \left[-4\ppnb+\ppng +2\left(1+\ppnz{2}+\ppnxi\right)\right]
    \right\},
    \notag 
    \\ & 
    \nu_3:=-2\cgw^2\ppng\left(1+\ppnz{3}\right),
    \quad 
    \nu_4:=-2\cgw^2\ppng \left[3\left(\ppng+\ppnz{4}\right)-2\ppnxi\right],
    \notag 
    \\ & 
    \nu_6:=\cgw^2\ppng\left(\ppnz{1}-2\ppnxi\right),
    \quad 
    \nu_W:=2\cgw^2\ppng\ppnxi,
    \notag 
    \\ & 
    \nu_{UU}:=-\frac{\beta_3}{4}+\frac{\cgw^2}{2}\left(-2+4\ppnb-5\ppng\right)\ppng 
    -2\delta_1,
    \notag 
    \\ & 
    \nu_{\mathcal{X}}:=\frac{3\beta_3}{2}+\ppng\left(3-2\cgw^2\ppngx-3\cgw^2\ppng\right)
        +\frac{3}{2\alpha_L}\left[
            2\beta_1+\beta_3+2\ppng\left(1-\cgw^2\ppng\right)
        \right],
\end{align}
and recall that $C$ is defined as Eq.~\eqref{C}.

Following the procedure described above,
the PPN parameters and $\ppngx$ are obtained as
\begin{align}
    \ppnb&=\frac{4\ppng\left[\cgw^2\ppng\left(1+\ppng\right)+2\delta_1\right]-\beta_3\left(3+\ppng\right)-2\delta_2}{4\left[2\cgw^2(\ppng)^2-\beta_3\right]},
    \\
    \ppna{2}&=\frac{3\left[2\left(\ppng+\beta_1\right)-2\cgw^2(\ppng)^2-\beta_3\right]^2}{2\left[2\cgw^2(\ppng)^2-\beta_3\right]}\left(\frac{1}{\alpha_L}+1\right)-1+\cgw^2(\ppng)^2+6\beta_1-\frac{\beta_3}{2}\notag\\
    &\quad+\frac{\beta_2-6\beta_1^2-12\beta_1\ppng}{2\cgw^2(\ppng)^2-\beta_3},
    \label{final-result-a2}
    \\ 
    \ppna{3}&=\ppnz{1}=\ppnz{2}=\ppnz{3}=\ppnz{4}=\ppnxi=0,
    \\ 
    \ppngx&=\frac{\beta_2+3\ppng\left[\beta_3+2\ppng\left(1-\cgw^2\ppng\right)\right]}%
    {2\left[2\cgw^2(\ppng)^2-2\beta_3\right]}
    +\frac{3\left(\beta_3+\ppng\right)\left[2\beta_1+\beta_3+2\ppng\left(1-\cgw^2\ppng\right)\right]}%
    {2\left[2\cgw^2(\ppng)^2-2\beta_3\right]\alpha_L},
    \label{final-result-gx}
\end{align}
while $d_{UU,W,1,2,3,4,6,\mathcal{X}}$ are found to be
\begin{align}
    &d_{UU}=\frac{3}{4}(\ppng)^2+\frac{\beta_3\left(3+2\ppng-4\ppnb\right)+2\delta_2}{8\cgw^2\ppng},
    \quad 
    d_W=d_6=0,
    \notag \\
    &d_1=\frac{\ppng}{2}+\frac{\beta_3}{2\cgw^2},
    \quad 
    d_2=-\frac{1}{2}(\ppng)^2-\frac{\beta_3\left(4\ppnb-3\right)+8\delta_1\ppng-2\delta_2}{4\cgw^2\ppng},
    \quad 
    d_3=\ppng,
    \notag \\ 
    &d_4=\frac{3\beta_3}{2\cgw^2},\quad 
    d_{\mathcal{X}}=\frac{\beta_2+2\ppngx\beta_3}{4\cgw^2\ppng}
    +\frac{3\left[2\beta_1+\beta_3+2\ppng\left(1-\cgw^2\ppng\right)\right]\beta_1}{4\cgw^2\ppng\alpha_L},
\end{align}
where $\ppnb$ and $\ppngx$ are used to shorten the expressions for $d_{UU}$, $d_2$, and $d_{\mathcal{X}}$.
In the above expressions we leave the EFT parameter $\beta_2$,
but one can replace it with $-6\beta_1^2/(1+\alpha_L)$ using the unitary degeneracy condition
(see Eq.~\eqref{eq:u-deg-condition}), which however does not simplify the equations much.
The EFT parameters $\delta_1$ and $\delta_2$, which have been introduced in Eqs.~\eqref{def:delta1} and~\eqref{def:delta2} and are not relevant to linear cosmology, appear only in $\ppnb$, the PPN parameter that measures nonlinearity in a superposition of gravity.
Note that the terms with the coefficients $d_{UU,W,1,2,3,4,6,\mathcal{X}}$ appear at $\mathcal{O}(\epsilon^4)$ in $g_{ij}$ and therefore are higher-order PN corrections, but they are necessary for determining the metric in the PN limit, as mentioned earlier.

Having thus determined all the PPN parameters,
we summarize our main results in Table~\ref{table:PPN-paras}
together with experimental constraints.

\renewcommand{\arraystretch}{2.2}
\begin{table}[tb]
\begin{tabular}{lccc}
\toprule
 Parameter & GR & Higher-order scalar-tensor theories & Constraints \\ 
 \midrule
 $\gn$ & $\ggw$ & $\displaystyle\left[\frac{2(1+\alpha_T)^2}{2(1+\alpha_H)^2-(1+\alpha_T)\beta_3}\right]\ggw$ & $0.995\lesssim \ggw/\gn\cgw\lesssim 1.00$~\cite{BeltranJimenez:2015sgd}$^a$\\
 \midrule
 $\cgw^2$ & 1 & $1+\alpha_T$ & $-3\times 10^{-15}<\cgw-1<7\times 10^{-16}$~\cite{LIGOScientific:2017vwq,LIGOScientific:2017zic}\\
 \midrule
 $\ppng$ & 1 & $\displaystyle{\frac{1+\alpha_H}{1+\alpha_T}}$ & $\ppng-1=(2.1\pm 2.3)\times 10^{-5}$~\cite{Bertotti:2003rm} \\
 $\ppnb$ & 1 & $\displaystyle\frac{4\ppng\left[\cgw^2\ppng\left(1+\ppng\right)+2\delta_1\right]-\beta_3\left(3+\ppng\right)-2\delta_2}{4\left[2\cgw^2(\ppng)^2-\beta_3\right]}$ & $\ppnb-1=(0.2\pm 2.5)\times 10^{-5}$~\cite{Verma:2013ata} \\
 $\ppnxi$ & 0 & 0 & $\checkmark$\\
 $\ppna{1}$ & 0 & $\displaystyle{4\left[2\cgw^2(\ppng)^2-\ppng-1-\beta_3\right]}$ & $\ppna{1}=-0.4^{+3.7}_{-3.1}\times 10^{-5}$~\cite{Shao:2012eg} \\
 $\ppna{2}$ & 0 & Eq.~\eqref{final-result-a2} & $|\ppna{2}|<1.6\times 10^{-9}$~\cite{Shao:2013wga}\\
 $\ppna{3}$ & 0 & 0 & $\checkmark$\\
 $\ppnz{1,2,3,4}$ & 0 & 0 & $\checkmark$ \\
 $\ppngx$ & N/A & Eq.~\eqref{final-result-gx}& N/A \\
\bottomrule
\multicolumn{4}{l}{\footnotesize{$^a$ The scalar gravitational radiation is ignored.}} \\
\end{tabular}
\caption{Summary of the PPN parameters. A checkmark indicates that the PPN parameter has the same value as in GR and hence experimental constraints are satisfied.}
\label{table:PPN-paras}
\end{table}

\section{Examples}\label{sec:examples}

In this section, we consider khronometric and TTDOF theories as examples.
In the former case, the PPN parameters were already determined in the literature~\cite{Blas:2010hb}.
Therefore, we can check the consistency of our general results.
In the latter case, we determine all the PPN parameters for the first time.

\subsection{Khronometric theory}
The action for khronometric theory~\cite{Blas:2009yd} is given by
\begin{align}
    S=\frac{M_*^2}{2}\int\D^4x\sqrt{-g}\left[
    \mathcal{R}+c_1\nabla_\mu u_\nu\nabla^\mu u^\nu+c_2(\nabla_\mu u^\mu)^2 
    +c_3\nabla_\mu u_\nu\nabla^\nu u^\mu+c_4u^\mu u^\nu \nabla_\mu u_\lambda \nabla_\nu
    u^\lambda
    \right],
\end{align}
where $u^\mu$ is a unit time-like vector field defined as $u_\mu:=-\phi_\mu/\sqrt{2X}$ and
$M_*$, $c_1$, $c_2$, $c_3$, and $c_4$ are constant parameters.
Since $\nabla_\mu u_\nu\nabla^\mu u^\nu$
can be expressed in terms of
$\nabla_\mu u_\nu\nabla^\nu u^\mu$ and $u^\mu u^\nu \nabla_\mu u_\lambda \nabla_\nu u^\lambda$,
one can set $c_1=0$ without loss of generality.
Then, it is straightforward to see that khronometric theory corresponds to
the scalar-tensor theory with
\begin{align}
    f=\frac{M_*^2}{2},\quad A_1=\frac{M_*^2}{2}\frac{c_3}{2X},\quad A_2=\frac{M_*^2}{2}\frac{c_2}{2X},\quad A_3=\frac{M_*^2}{2}\frac{c_2}{2X^2}, \quad 
    \quad A_4=\frac{M_*^2}{2}\frac{2c_3+c_4}{4X^2},\quad A_5=\frac{M_*^2}{2}\frac{c_3+c_4}{8X^2}.
\end{align}
We thus have~\cite{Langlois:2017mxy}
\begin{align}
    M^2=(1+c_3)M_*^2,\quad\alpha_T=\alpha_H=-\frac{c_3}{1+c_3},\quad \alpha_L=-\frac{3}{2}\frac{c_2+c_3}{1+c_3},\quad \beta_1=\beta_2=0,\quad \beta_3=\frac{c_4}{1+c_3},\quad \delta_1=\delta_2=0,
\end{align}
from which we immediately get
\begin{align}
8\pi \gn&=\frac{1}{(1-c_4/2)M_*^2},\quad 
\ppng=1,\quad \beta=1,\quad\alpha_1=-\frac{4(2c_3+c_4)}{1+c_3},
\notag \\
\ppna{2}&=-2+\frac{4}{1+c_3}-\frac{2c_2}{c_2+c_3}-\frac{3(2-3c_2-c_3)}{3c_2+c_3-c_4}+\frac{(1-2c_2-c_3)c_4}{(1+c_3)(c_2+c_3)},
\end{align}
with $\ppna{3}=\ppnz{1}=\ppnz{2}=\ppnz{3}=\ppnz{4}=\ppnxi=0$.
This correctly reproduces the previous result obtained in the literature~\cite{Blas:2010hb, Blas:2011zd}.

\subsection{TTDOF theory}\label{subsec:ttdof}

Let us calculate the PPN parameters in theories with two tensor degrees of freedom and no scalar propagating mode.
The scalar field thus is a generalization of a cuscuton field~\cite{Afshordi:2006ad},
yielding an instantaneous mode.
A family of such theories was proposed in Ref.~\cite{Gao:2019twq} (see also~\cite{Hu:2021yaq}).
The action was originally given in terms of the Arnowitt-Deser-Misner variables
in the unitary gauge~\cite{Gao:2019twq}, and later it was written in a fully covariant form as a limiting case of the U-DHOST theory~\cite{Iyonaga:2021yfv}.
The functions in the action of the shift-symmetric TTDOF theory
we consider are given by
\begin{align}
    f&=\frac{1}{2}\left(\bar a_2+\bar a_4\sqrt{2X}\right),\quad A_1=\frac{1}{4X}\left(\frac{\bar b_0}{1+\bar b_2\sqrt{2X}}\right)-\frac{f}{2X},
    \quad 
    A_2=-A_1+4X^2A_5,
    \notag \\ 
    A_3&=-A_4+4XA_5,
    \quad 
    A_4=\frac{f'+A_1}{X},\quad A_5=\frac{1}{24X^3}\cdot\frac{\bar b_0\left(\bar b_1-\bar b_2\right)\sqrt{2X}}{\left(1+\bar b_1\sqrt{2X}\right)\left(1+\bar b_2\sqrt{2X}\right)},
\label{Coeff: TTDOF}
\end{align}
where $\bar a_2$, $\bar a_4$, $\bar b_0$, $\bar b_1$, and $\bar b_2$ are constant parameters of the theory.
In the case of $\bar b_1=\bar b_2$, this reduces to the extended cuscuton theory~\cite{Iyonaga:2018vnu}.

The EFT parameters are found to be
\begin{align}
    M^2&=\frac{\bar b_0}{1+\bar b_2q},\quad M^2(1+\alpha_T)=\bar a_2+\bar a_4q,\quad M^2(1+\alpha_H)=\bar a_2,\quad \alpha_L=\frac{\left(\bar b_2-\bar b_1\right)q}{1+\bar b_1q},
    \notag \\ 
    \beta_1&=\beta_2=\beta_3=\delta_1=\delta_2=0,
\end{align}
where in the present case $\dot\phi=q$ is an arbitrary constant.
The Newtonian gravitational constant, the speed of gravitational waves, and
the nontrivial PPN parameters are then given by
\begin{align}
    8\pi \gn&=\frac{\bar a_2+\bar a_4q}{\bar a_2^2},
    \quad 
    \cgw^2=\frac{\left(\bar a_2+\bar a_4q\right)\left(1+\bar b_2q\right)}{\bar b_0},
    \quad   
    \ppng=\frac{\bar a_2}{\bar a_2+\bar a_4q},\quad \ppnb=\frac{1+\ppng}{2},
    \notag \\  
    \ppna{1}&=4\left[2\cgw^2(\ppng)^2-\ppng-1\right],
    \quad 
    \ppna{2}=\frac{3\left(1-\cgw^2\ppng\right)^2}{\cgw^2}\left(\frac{1}{\alpha_L}+1\right)-1+\cgw^2(\ppng)^2.
\end{align}
One can see from this that
\begin{align}
    \bar a_2=\frac{\bar b_0}{1+\bar b_2q},\quad \bar a_4=0
    \quad \Leftrightarrow
    \quad 
    \gn=\ggw,\quad \cgw=1,\quad \ppng=\ppnb=1,
    \quad 
    \ppna{1}=\ppna{2}=0,
\end{align}
regardless of the value of $\alpha_L$.
Therefore, even though the theory is not equivalent to GR at the level of the action,
the PPN parameters in this case are equal to the values in GR
and gravitational waves propagate at the speed of light.
The presence of such TTDOF theories has been suggested in Ref.~\cite{Iyonaga:2021yfv}
by evaluating only $\cgw$ and $\ppng$.
Here we have found that all the other PPN parameters agree with the values in GR as well.

\section{Experimental constraints}\label{sec:constraints_EFT para}

\subsection{Constraints on the EFT parameters}

  \begin{figure}[tb]
    \begin{center}
        \includegraphics[keepaspectratio=true,height=100mm]{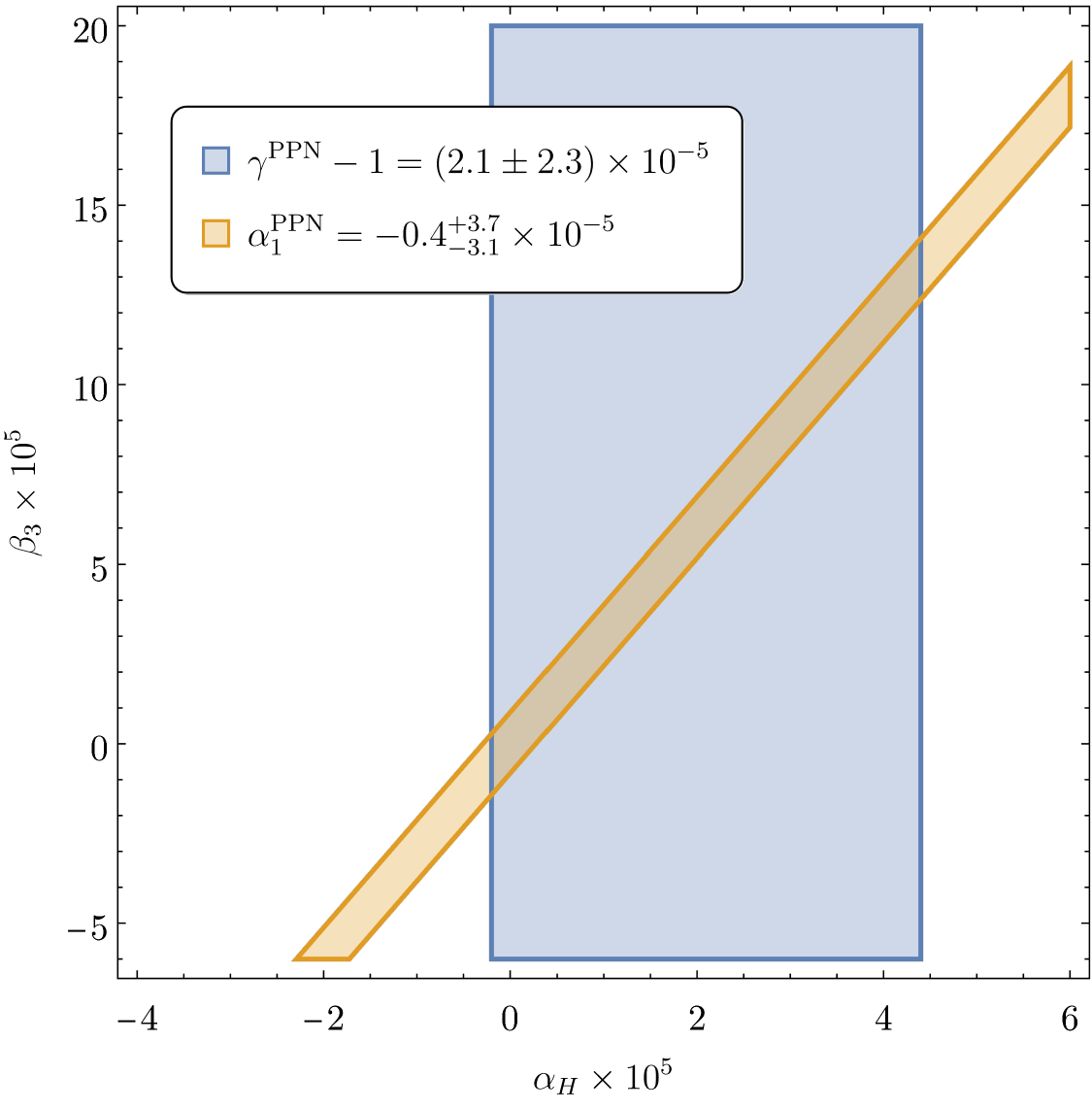}
    \end{center}
        \caption{Constraints on $\alpha_H$ and $\beta_3$ derived from the experimental bounds on the PPN parameters $\ppng$~\cite{Bertotti:2003rm} and $\ppna{1}$~\cite{Shao:2012eg}.
        The thin parallelogram embodied as the overlapping region of the two constraints indicates the allowed parameter space.}
    \label{fig:constraint-on-ppn.pdf}
  \end{figure}

  \begin{figure}[tb]
    \begin{center}
        \includegraphics[keepaspectratio=true,height=100mm]{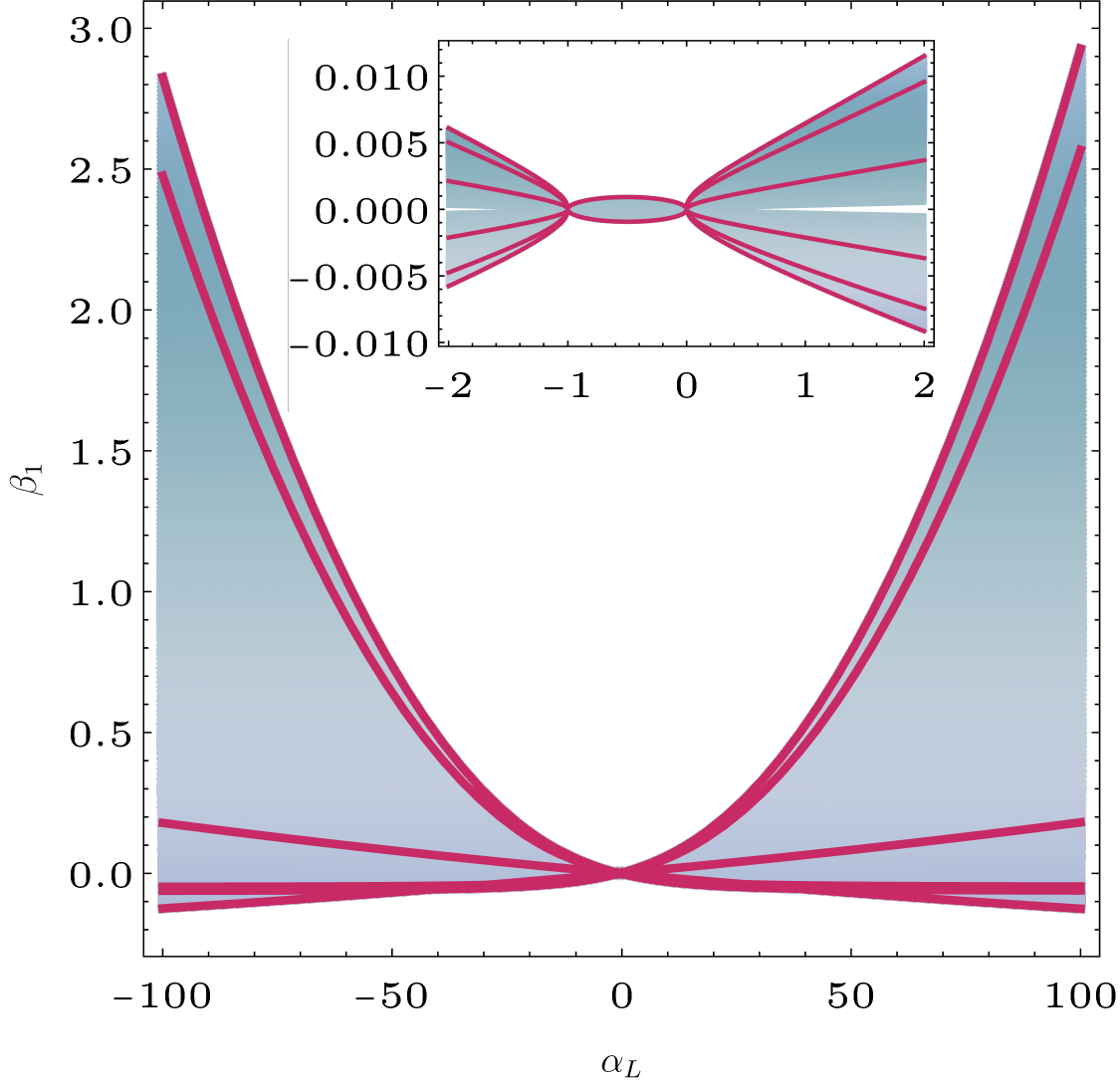}
    \end{center}
        \caption{The region satisfying $\ppna{2}=0$ for the experimentally viable parameter range of $(\alpha_H,\beta_3)$ is shown shaded.
        The magenta curves are drawn with $(\alpha_H,\beta_3)$ at the four vertices of the thin parallelogram in Fig.~\ref{fig:constraint-on-ppn.pdf}.
        The inset shows a zoom of the region near the origin.}
    \label{fig:aLb1.pdf}
  \end{figure}

In the previous section, we determined the PPN parameters in terms of the EFT of dark energy parameters under the assumption that $\alpha_L \ne 0$.
Let us derive constraints on the EFT parameters from experimental constraints on the PPN parameters.
To do so, we assume that the propagation speed of gravitational waves is equal to the speed of light because it is particularly tightly constrained by the observations of GW170817~\cite{LIGOScientific:2017vwq} and GRB170817A~\cite{LIGOScientific:2017zic}.
Setting $\cgw=1$, the Newtonian gravitational constant and the PPN parameters read
\begin{align}
    \gn&=\frac{\ggw}{(1+\alpha_H)^2-\beta_3/2}\label{N-vs-GW},
    \\ 
    \ppng&=1+\alpha_H,
    \\
    \ppnb&=\frac{4\ppng\left[\ppng\left(1+\ppng\right)+2\delta_1\right]-\beta_3\left(3+\ppng\right)-2\delta_2}{4\left[2(\ppng)^2-\beta_3\right]},\\
    \ppna{1}&=4\left[2(\ppng)^2-\ppng-1-\beta_3\right],\\
    \ppna{2}&=\frac{3\left[2(\ppng+\beta_1)-2(\ppng)^2-\beta_3\right]^2}{2\left[2(\ppng)^2-\beta_3\right]}\left(\frac{1}{\alpha_L}+1\right)-1+(\ppng)^2+6\beta_1-\frac{\beta_3}{2}
    +\frac{\beta_2-6\beta_1^2-12\beta_1\ppng}{2(\ppng)^2-\beta_3}.
\end{align}

Experimental constraints on the PPN parameters are summarized in~\cite{Will:2018bme} and are replicated for the convenience of a reader in Table~\ref{table:PPN-paras}.
Since $\ppng$ and $\ppna{1}$ are written solely in terms of $\alpha_H$ and $\beta_3$, the constraints on the former parameters can immediately be translated to the latter.
The constraints on $\alpha_H$ and $\beta_3$ thus derived are shown in Fig.~\ref{fig:constraint-on-ppn.pdf}.

Equation~\eqref{N-vs-GW} implies that the effective gravitational coupling for gravitational waves differs from the Newtonian gravitational constant
and the deviation is controlled also by $\alpha_H$ and $\beta_3$.
The effective gravitational coupling for gravitational waves can be measured e.g. through the energy loss from a binary system provided that it is dominated by quadrupole radiation of usual tensorial gravitational waves.
The authors of Ref.~\cite{BeltranJimenez:2015sgd} thus obtained the bound 
$0.995\lesssim \ggw/\gn\cgw\lesssim 1.00$
at the 1$\sigma$ level. This bound is weaker than those on $\ppng$ and $\ppna{1}$.
Note, however, that in the present case it is unclear whether or not the assumption of the absence of monopole and dipole radiation is justified.
Evaluating the scalar gravitational radiation from a binary system in U-DHOST theories is beyond the scope of the paper, but is an interesting future direction.

Having obtained the constraints on $\alpha_H$ and $\beta_3$,
we now assume that $|\alpha_H|\ll 1$ and $|\beta_3|\ll 1$ and write $\ppnb$ as
$\ppnb\simeq 1+\delta_1-\delta_2/4+\alpha_H/2$.
Then, the experimental bound on $\ppnb$ can be translated to the constraint on
the particular combination of the EFT parameters, including the ones newly introduced in the present paper:
\begin{align}
    \delta_1
    -\frac{\delta_2}{4}+\frac{\alpha_H}{2}=(0.2\pm 2.5)\times 10^{-5}.
\end{align}

Upon imposing the unitary degeneracy condition~\eqref{eq:u-deg-condition},
$\ppna{2}$ depends on the four EFT parameters, $\alpha_H$, $\beta_3$, $\alpha_L$, and $\beta_1$, in a complicated way.
The errors in $\alpha_H$ and $\beta_3$ are of $\mathcal{O}(10^{-5})$,
while the experimental bound on $\ppna{2}$ is as small as one part in $10^9$~\cite{Shao:2013wga}.
Therefore, we present in Fig.~\ref{fig:aLb1.pdf} the region in the $\alpha_L$\nobreakdash--$\beta_1$~plane satisfying $\ppna{2}=0$ for the experimentally viable range of $\alpha_H$ and $\beta_3$.
It can be seen that a large parameter region in the $\alpha_L$\nobreakdash--$\beta_1$~plane is still allowed.

\subsection{U-DHOST theory that is indistinguishable from GR}

It is interesting to explore higher-order scalar-tensor theories that are indistinguishable from GR at the level of the PPN parameters and thus satisfy all experimental constraints on them trivially.
The discussion in Sec.~\ref{subsec:ttdof} shows that such a theory does exist.
We now determine the general form of the Lagrangian for the theories having such properties.
To do so, we first set
\begin{align}
    \alpha_T=\alpha_H=\beta_3=0,\label{TH3=0}
\end{align}
so that $\cgw=1$, $\ppng=1$, and $\ppna{1}=0$.
In terms of the functions in the Lagrangian, Eq.~\eqref{TH3=0} reads
\begin{align}
    f'=A_1=A_4=0.\label{TH3=0-2}
\end{align}
Strictly speaking, these conditions are required to be satisfied only at $X=q^2/2$.
However, let us require Eq.~\eqref{TH3=0-2} to hold at any $X$.
Then, we have $\delta_1=\delta_2=0$, and hence $\ppnb=1$.
Finally, using the unitary degeneracy condition~\eqref{eq:u-deg-condition},
we obtain
\begin{align}
    \ppna{2}=\frac{3\beta_1^2}{\alpha_L}
    +\frac{\beta_2}{2}=\frac{3\beta_1^2}{\alpha_L(1+\alpha_L)}.
\end{align}
We therefore require that 
\begin{align}
    \beta_1=0\quad\Rightarrow\quad A_3=\frac{A_2}{X}.\label{cond-A3}
\end{align}
The unitary degeneracy condition~\eqref{eq:u-deg-condition} itself leads to
\begin{align}
    A_5=\frac{A_2}{4X^2}.\label{cond-A5}
\end{align}
The U-DHOST theory satisfying Eqs.~\eqref{TH3=0-2},~\eqref{cond-A3}, and~\eqref{cond-A5}
and characterized by a single function $A_2(X)$ yields the same PPN parameters as in GR irrespective of the value of $\alpha_L=-3XA_2/f|_{X=q^2/2}$.

The Lagrangian for the U-DHOST theory that is indistinguishable from GR at the level of the PPN parameters takes the following simple form:
\begin{align}
    \mathcal{L}=\frac{\mpl^2}{2}\left[ 
    \mathcal{R}-\frac{\alpha(X)}{3X}\left(\Box\phi+\frac{\phi^\mu\phi_{\mu\nu}\phi^\nu}{2X}\right)^2
    \right]+P(X),
\end{align}
where $\mpl\,(:=\sqrt{2f})$ is the Planck mass and $\alpha(X)\,(:=-3XA_2/f)$ is an arbitrary function of $X$.
It should be noted that in the unitary gauge, the term proportional to $\alpha(X)$ reduces to the square of the trace of the extrinsic curvature of constant time hypersurfaces.\footnote{Very recently, a similar subset of Einstein-aether theory has attracted some interest~\cite{Franzin:2023rdl,Gurses:2024jkx}.}
It would be interesting to test this restricted class of U-DHOST theories with cosmological probes.

\section{Conclusions}\label{conclusions}

In this paper, we have studied the post-Newtonian (PN) limit of ghost-free higher-order scalar-tensor theories and determined all the parametrized post-Newtonian (PPN) parameters.
The Lagrangian of scalar-tensor theories we considered contains general quadratic terms built out of second derivatives of the scalar field~\cite{Langlois:2015cwa}.
Ghost instabilities due to this higher-derivative nature can be avoided by imposing the unitary degeneracy condition, yielding the unitary degenerate higher-order scalar-tensor (U-DHOST) family of scalar-tensor theories~\cite{DeFelice:2018ewo}, though the condition itself plays no particular role in the PN calculations.
Higher-order scalar-tensor theories can be incorporated into the framework of the effective field theory (EFT) of dark energy and thus be conveniently characterized by several parameters of the EFT of dark energy~\cite{Gubitosi:2012hu,Gleyzes:2013ooa,Bellini:2014fua,Langlois:2017mxy}, which have been frequently used to discuss cosmological constraints on scalar-tensor theories and dark energy models.
We have shown that the four PPN parameters $\ppng$, $\ppnb$, $\ppna{1}$, and $\ppna{2}$ can deviate from the values in general relativity (GR) and given their explicit expressions in terms of the EFT of dark energy parameters.
The other PPN parameters have been shown to take the same values as in GR.
Our main results are summarized in Table~\ref{table:PPN-paras}.
The crucial assumptions made in our analysis are that the gradient of the scalar field is timelike and the EFT parameter $\alpha_L$ is nonvanishing, which is typically the case in Lorentz-breaking theories.
We have checked that our general results correctly reproduce the previous ones for the khronometric theory in the appropriate limit~\cite{Blas:2010hb,Blas:2011zd}.

Using the PPN parameters expressed in terms of the EFT parameters, we have put the experimental constraints on the latter parameters.
Under the assumption that gravitational waves propagate exactly at the speed of light, we have found that the bounds on the EFT parameters $\alpha_H$ and $\beta_3$ are at the level of $\mathcal{O}(10^{-5})$, while a large parameter space is still allowed for $\alpha_L$ and $\beta_1$.
Furthermore, we have explored scalar-tensor theories all of whose PPN parameters have the same values as in GR, and identified the form of the Lagrangian for such theories.
It has been found that the Lagrangian is characterized by a single function of the kinetic term of the scalar field, $X=-(\partial\phi)^2/2$, (aside from the $k$-essence part of the Lagrangian) and there remains a nonvanishing EFT parameter $\alpha_L$, which can be arbitrary as far as the PN limit is concerned.

It would be interesting to study gravitational radiation from PN sources, which would provide novel tests of the U-DHOST family of scalar-tensor theories.
Combining the PN tests with cosmological observations would offer further constraints on the EFT of dark energy parameters.
These are the future directions which we believe are worth pursuing.

\acknowledgments
The work of JS was supported by
the Rikkyo University Special Fund for Research.
The work of TK was supported by
JSPS KAKENHI Grant No.~JP20K03936 and
MEXT-JSPS Grant-in-Aid for Transformative Research Areas (A) ``Extreme Universe'',
No.~JP21H05182 and No.~JP21H05189.

\appendix

\section{On the ansatz for the scalar field and the assumption $\alpha_L\neq 0$}\label{app:aL}

Let us see what happens if the scalar field has a $\mathcal{O}(\epsilon^2)$ contribution:
\begin{align}
    \phi=qt+\delta\phi^{(2)}(t,\Vec{x})+\mathcal{O}(\epsilon^3)
    ,\quad \delta\phi^{(2)}=\mathcal{O}(\epsilon^2).
\end{align}
The $(0i)$ component of the field equations then starts at $\mathcal{O}(\epsilon^2)$:
\begin{align}
    \mathcal{E}_{0i}=-2q(A_1+A_2)\partial_i\Delta\delta\phi^{(2)}+\mathcal{O}(\epsilon^3).
\end{align}
This shows that we have $\delta\phi^{(2)}=0$ as long as $\alpha_L=-(6X/M^2)(A_1+A_2)\neq 0$,
which justifies Eq.~\eqref{PN-ex-phi}.
This also clarifies why we need to assume $\alpha_L\neq 0$ for the validity of our analysis.
In theories with $\alpha_L=0$, the scalar field may have an $\mathcal{O}(\epsilon^2)$ correction.

\section{Gravitational field equations}\label{app:A}

In this appendix, we present field equations derived from the action~\eqref{eq:U-DHOST-action} with shift-symmetry, i.e. in the case where the coefficients are considered as the functions of $X:=-g^{\mu\nu}\phi_\mu\phi_\nu/2$ only.
By taking the variations of Eq.~\eqref{eq:U-DHOST-action} with respect to the metric and scalar field respectively, we obtain the field equations as
\begin{align}
\mathcal{E}_{\mu\nu}:=\frac{2}{\sqrt{-g}}\frac{\delta S_{\text{grav}}}{\delta g^{\mu\nu}}=\mathcal{E}^{(P)}_{\mu\nu}+\mathcal{E}^{(f)}_{\mu\nu}+\sum_{I=1}^{5}\mathcal{E}^{(I)}_{\mu\nu},
\quad
\mathcal{E}_\phi:=\frac{1}{\sqrt{-g}}\frac{\delta S_{\text{grav}}}{\delta \phi},
\end{align}
where we find
\begin{align}
\mathcal{E}_{\mu\nu}^{\left(P\right)} &:= -g_{\mu\nu}P-\phi_{\mu}\phi_{\nu}P^{\prime},\\
\mathcal{E}_{\mu\nu}^{\left(f\right)} &:= 2G_{\mu\nu}f+\mathcal{E}_{\mu\nu}^{\left(f^{\prime}\right)}f^{\prime}+\mathcal{E}_{\mu\nu}^{\left(f^{\prime\prime}\right)}f^{\prime\prime},\\
\mathcal{E}_{\mu\nu}^{\left(I\right)} &:= \mathcal{E}_{\mu\nu}^{\left(A_{I}\right)}A_{I}+\mathcal{E}_{\mu\nu}^{\left(A_{I}^{\prime}\right)}A_{I}^{\prime},
\end{align}
with the expressions
\begin{align}
\mathcal{E}_{\mu\nu}^{\left(f^{\prime}\right)} & :=  -\mathcal{R}\phi_{\mu}\phi_{\nu}-2g_{\mu\nu}\left(\phi^{\alpha}{}_{\alpha\beta}\phi^{\beta}+\phi_{\alpha\beta}\phi^{\alpha\beta}\right)+2\left(\phi_{\alpha\mu}\phi_{\nu}{}^{\alpha}+\phi_{(\mu\nu)\alpha}\phi^{\alpha}\right),\\
\mathcal{E}_{\mu\nu}^{\left(f^{\prime\prime}\right)} & :=  2\left(g_{\mu\nu}\phi_{\alpha}{}^{\rho}\phi_{\rho\beta}-\phi_{\mu\alpha}\phi_{\nu\beta}\right)\phi^{\alpha}\phi^{\beta},\\
\mathcal{E}_{\mu\nu}^{\left(A_{1}\right)} & :=  -g_{\mu\nu}\phi_{\alpha\beta}\phi^{\alpha\beta}-4\phi^{\alpha}{}_{\alpha(\mu}\phi_{\nu)}+2\left(\phi_{\mu\nu}\phi_{\alpha}{}^{\alpha}+\phi_{\alpha\mu\nu}\phi^{\alpha}\right),\\
\mathcal{E}_{\mu\nu}^{\left(A_{1}^{\prime}\right)} & :=  \left(4\phi_{(\mu}\phi_{\nu)\alpha}\phi_{\beta}-2\phi_{\mu\nu}\phi_{\alpha}\phi_{\beta}-\phi_{\mu}\phi_{\nu}\phi_{\alpha\beta}\right)\phi^{\alpha\beta},\\
\mathcal{E}_{\mu\nu}^{\left(A_{2}\right)} & :=  g_{\mu\nu}\left(2\phi^{\alpha}\phi_{\alpha\beta}{}^{\beta}+\phi_{\alpha}{}^{\alpha}\phi_{\beta}{}^{\beta}\right)-4\phi_{(\mu}\phi_{\nu)\alpha}{}^{\alpha},\\
\mathcal{E}_{\mu\nu}^{\left(A_{2}^{\prime}\right)} & :=  -2g_{\mu\nu}\phi_{\alpha\beta}\phi^{\alpha}\phi^{\beta}\phi_{\rho}{}^{\rho}-\phi_{\mu}\phi_{\nu}\phi_{\alpha}{}^{\alpha}\phi_{\beta}{}^{\beta}+4\phi_{(\mu}\phi_{\nu)\alpha}\phi^{\alpha}\phi_{\beta}{}^{\beta},\\
\mathcal{E}_{\mu\nu}^{\left(A_{3}\right)} & :=  g_{\mu\nu}\left(\phi^{\rho}\phi_{\rho\beta\alpha}+2\phi_{\alpha}{}^{\rho}\phi_{\rho\beta}\right)\phi^{\alpha}\phi^{\beta}-\phi_{\mu}\phi_{\nu}\left(\phi^{\alpha}\phi_{\alpha\beta}{}^{\beta}+\phi_{\alpha}{}^{\alpha}\phi_{\beta}{}^{\beta}\right)\notag \\
    & \quad-4\phi_{(\mu}\phi_{\nu)\alpha}\phi_{\beta}\phi^{\alpha\beta}+2\phi_{(\mu}\phi_{\nu)\alpha}\phi^{\alpha}\phi_{\beta}{}^{\beta}-2\phi_{(\mu}\phi_{\nu)\alpha\beta}\phi^{\alpha}\phi^{\beta},\\
\mathcal{E}_{\mu\nu}^{\left(A_{3}^{\prime}\right)} & :=  \left(-g_{\mu\nu}\phi_{\alpha\beta}\phi_{\rho\sigma}\phi^{\sigma}+2\phi_{(\mu}\phi_{\nu)\alpha}\phi_{\beta\rho}\right)\phi^{\alpha}\phi^{\beta}\phi^{\rho},\\
\mathcal{E}_{\mu\nu}^{\left(A_{4}\right)} & :=  \left(-g_{\mu\nu}\phi_{\alpha}{}^{\rho}\phi_{\rho\beta}+2\phi_{\mu\alpha}\phi_{\nu\beta}\right)\phi^{\alpha}\phi^{\beta}-2\phi_{\mu}\phi_{\nu}\left(\phi^{\alpha}{}_{\alpha\beta}\phi^{\beta}+\phi_{\alpha\beta}\phi^{\alpha\beta}\right),\\
\mathcal{E}_{\mu\nu}^{\left(A_{4}^{\prime}\right)} & :=  \phi_{\mu}\phi_{\nu}\phi^{\alpha}\phi_{\alpha}{}^{\rho}\phi_{\rho\beta}\phi^{\beta},\\
\mathcal{E}_{\mu\nu}^{\left(A_{5}\right)} & :=  -g_{\mu\nu}\phi_{\alpha\beta}\phi^{\alpha}\phi^{\beta}\phi_{\rho\sigma}\phi^{\rho}\phi^{\sigma}+4\phi_{(\mu}\phi_{\nu)\alpha}\phi_{\beta\rho}\phi^{\alpha}\phi^{\beta}\phi^{\rho}
\notag \\ &\quad 
  -2\phi_{\mu}\phi_{\nu}\left(\phi_{\alpha\beta}\phi_{\rho}{}^{\rho}+\phi^{\rho}\phi_{\rho\beta\alpha}+2\phi_{\alpha}{}^{\rho}\phi_{\rho\beta}\right)\phi^{\alpha}\phi^{\beta},\\
\mathcal{E}_{\mu\nu}^{\left(A_{5}^{\prime}\right)} & :=  \phi_{\mu}\phi_{\nu}\phi_{\alpha\beta}\phi^{\alpha}\phi^{\beta}\phi_{\rho\sigma}\phi^{\rho}\phi^{\sigma},
\end{align}
and we denote $\phi_{\alpha\mu\nu}:=\nabla_{\alpha}\nabla_{\mu}\nabla_{\nu}\phi$.
The Bianchi identity,
\begin{align}
    \nabla^\nu\mathcal{E}_{\mu\nu}=-\phi_\mu \mathcal{E}_\phi,
\end{align}
follows from the general covariance of the theory.
We therefore do not present the explicit form of $\mathcal{E}_\phi$,
which is not used in the main text.


\bibliography{refs}

\providecommand{\href}[2]{#2}\begingroup\raggedright\begin{thebibliography}{10}

\bibitem{Clifton:2011jh}
T.~Clifton, P.~G. Ferreira, A.~Padilla and C.~Skordis, \emph{{Modified Gravity and Cosmology}}, \href{https://doi.org/10.1016/j.physrep.2012.01.001}{\emph{Phys. Rept.} {\bfseries 513} (2012) 1} [\href{https://arxiv.org/abs/1106.2476}{{\ttfamily 1106.2476}}].

\bibitem{Heisenberg:2018vsk}
L.~Heisenberg, \emph{{A systematic approach to generalisations of General Relativity and their cosmological implications}}, \href{https://doi.org/10.1016/j.physrep.2018.11.006}{\emph{Phys. Rept.} {\bfseries 796} (2019) 1} [\href{https://arxiv.org/abs/1807.01725}{{\ttfamily 1807.01725}}].

\bibitem{Horndeski:1974wa}
G.~W. Horndeski, \emph{{Second-order scalar-tensor field equations in a four-dimensional space}}, \href{https://doi.org/10.1007/BF01807638}{\emph{Int. J. Theor. Phys.} {\bfseries 10} (1974) 363}.

\bibitem{Deffayet:2011gz}
C.~Deffayet, X.~Gao, D.~A. Steer and G.~Zahariade, \emph{{From k-essence to generalised Galileons}}, \href{https://doi.org/10.1103/PhysRevD.84.064039}{\emph{Phys. Rev. D} {\bfseries 84} (2011) 064039} [\href{https://arxiv.org/abs/1103.3260}{{\ttfamily 1103.3260}}].

\bibitem{Kobayashi:2011nu}
T.~Kobayashi, M.~Yamaguchi and J.~Yokoyama, \emph{{Generalized G-inflation: Inflation with the most general second-order field equations}}, \href{https://doi.org/10.1143/PTP.126.511}{\emph{Prog. Theor. Phys.} {\bfseries 126} (2011) 511} [\href{https://arxiv.org/abs/1105.5723}{{\ttfamily 1105.5723}}].

\bibitem{Woodard:2006nt}
R.~P. Woodard, \emph{{Avoiding dark energy with 1/r modifications of gravity}}, \href{https://doi.org/10.1007/978-3-540-71013-4_14}{\emph{Lect. Notes Phys.} {\bfseries 720} (2007) 403} [\href{https://arxiv.org/abs/astro-ph/0601672}{{\ttfamily astro-ph/0601672}}].

\bibitem{Zumalacarregui:2013pma}
M.~Zumalac\'arregui and J.~Garc\'\i{}a-Bellido, \emph{{Transforming gravity: from derivative couplings to matter to second-order scalar-tensor theories beyond the Horndeski Lagrangian}}, \href{https://doi.org/10.1103/PhysRevD.89.064046}{\emph{Phys. Rev. D} {\bfseries 89} (2014) 064046} [\href{https://arxiv.org/abs/1308.4685}{{\ttfamily 1308.4685}}].

\bibitem{Gleyzes:2014dya}
J.~Gleyzes, D.~Langlois, F.~Piazza and F.~Vernizzi, \emph{{Healthy theories beyond Horndeski}}, \href{https://doi.org/10.1103/PhysRevLett.114.211101}{\emph{Phys. Rev. Lett.} {\bfseries 114} (2015) 211101} [\href{https://arxiv.org/abs/1404.6495}{{\ttfamily 1404.6495}}].

\bibitem{Langlois:2015cwa}
D.~Langlois and K.~Noui, \emph{{Degenerate higher derivative theories beyond Horndeski: evading the Ostrogradski instability}}, \href{https://doi.org/10.1088/1475-7516/2016/02/034}{\emph{JCAP} {\bfseries 02} (2016) 034} [\href{https://arxiv.org/abs/1510.06930}{{\ttfamily 1510.06930}}].

\bibitem{Crisostomi:2016czh}
M.~Crisostomi, K.~Koyama and G.~Tasinato, \emph{{Extended Scalar-Tensor Theories of Gravity}}, \href{https://doi.org/10.1088/1475-7516/2016/04/044}{\emph{JCAP} {\bfseries 04} (2016) 044} [\href{https://arxiv.org/abs/1602.03119}{{\ttfamily 1602.03119}}].

\bibitem{BenAchour:2016fzp}
J.~Ben~Achour, M.~Crisostomi, K.~Koyama, D.~Langlois, K.~Noui and G.~Tasinato, \emph{{Degenerate higher order scalar-tensor theories beyond Horndeski up to cubic order}}, \href{https://doi.org/10.1007/JHEP12(2016)100}{\emph{JHEP} {\bfseries 12} (2016) 100} [\href{https://arxiv.org/abs/1608.08135}{{\ttfamily 1608.08135}}].

\bibitem{BenAchour:2016cay}
J.~Ben~Achour, D.~Langlois and K.~Noui, \emph{{Degenerate higher order scalar-tensor theories beyond Horndeski and disformal transformations}}, \href{https://doi.org/10.1103/PhysRevD.93.124005}{\emph{Phys. Rev. D} {\bfseries 93} (2016) 124005} [\href{https://arxiv.org/abs/1602.08398}{{\ttfamily 1602.08398}}].

\bibitem{Langlois:2018dxi}
D.~Langlois, \emph{{Dark energy and modified gravity in degenerate higher-order scalar\textendash{}tensor (DHOST) theories: A review}}, \href{https://doi.org/10.1142/S0218271819420069}{\emph{Int. J. Mod. Phys. D} {\bfseries 28} (2019) 1942006} [\href{https://arxiv.org/abs/1811.06271}{{\ttfamily 1811.06271}}].

\bibitem{Kobayashi:2019hrl}
T.~Kobayashi, \emph{{Horndeski theory and beyond: a review}}, \href{https://doi.org/10.1088/1361-6633/ab2429}{\emph{Rept. Prog. Phys.} {\bfseries 82} (2019) 086901} [\href{https://arxiv.org/abs/1901.07183}{{\ttfamily 1901.07183}}].

\bibitem{DeFelice:2018ewo}
A.~De~Felice, D.~Langlois, S.~Mukohyama, K.~Noui and A.~Wang, \emph{{Generalized instantaneous modes in higher-order scalar-tensor theories}}, \href{https://doi.org/10.1103/PhysRevD.98.084024}{\emph{Phys. Rev. D} {\bfseries 98} (2018) 084024} [\href{https://arxiv.org/abs/1803.06241}{{\ttfamily 1803.06241}}].

\bibitem{DeFelice:2021hps}
A.~De~Felice, S.~Mukohyama and K.~Takahashi, \emph{{Nonlinear definition of the shadowy mode in higher-order scalar-tensor theories}}, \href{https://doi.org/10.1088/1475-7516/2021/12/020}{\emph{JCAP} {\bfseries 12} (2021) 020} [\href{https://arxiv.org/abs/2110.03194}{{\ttfamily 2110.03194}}].

\bibitem{Gao:2014soa}
X.~Gao, \emph{{Unifying framework for scalar-tensor theories of gravity}}, \href{https://doi.org/10.1103/PhysRevD.90.081501}{\emph{Phys. Rev. D} {\bfseries 90} (2014) 081501} [\href{https://arxiv.org/abs/1406.0822}{{\ttfamily 1406.0822}}].

\bibitem{Gao:2014fra}
X.~Gao, \emph{{Hamiltonian analysis of spatially covariant gravity}}, \href{https://doi.org/10.1103/PhysRevD.90.104033}{\emph{Phys. Rev. D} {\bfseries 90} (2014) 104033} [\href{https://arxiv.org/abs/1409.6708}{{\ttfamily 1409.6708}}].

\bibitem{Gao:2018znj}
X.~Gao and Z.-B. Yao, \emph{{Spatially covariant gravity with velocity of the lapse function: the Hamiltonian analysis}}, \href{https://doi.org/10.1088/1475-7516/2019/05/024}{\emph{JCAP} {\bfseries 05} (2019) 024} [\href{https://arxiv.org/abs/1806.02811}{{\ttfamily 1806.02811}}].

\bibitem{Arkani-Hamed:2003pdi}
N.~Arkani-Hamed, H.-C. Cheng, M.~A. Luty and S.~Mukohyama, \emph{{Ghost condensation and a consistent infrared modification of gravity}}, \href{https://doi.org/10.1088/1126-6708/2004/05/074}{\emph{JHEP} {\bfseries 05} (2004) 074} [\href{https://arxiv.org/abs/hep-th/0312099}{{\ttfamily hep-th/0312099}}].

\bibitem{Horava:2008ih}
P.~Horava, \emph{{Membranes at Quantum Criticality}}, \href{https://doi.org/10.1088/1126-6708/2009/03/020}{\emph{JHEP} {\bfseries 03} (2009) 020} [\href{https://arxiv.org/abs/0812.4287}{{\ttfamily 0812.4287}}].

\bibitem{Horava:2009uw}
P.~Horava, \emph{{Quantum Gravity at a Lifshitz Point}}, \href{https://doi.org/10.1103/PhysRevD.79.084008}{\emph{Phys. Rev. D} {\bfseries 79} (2009) 084008} [\href{https://arxiv.org/abs/0901.3775}{{\ttfamily 0901.3775}}].

\bibitem{Blas:2009yd}
D.~Blas, O.~Pujolas and S.~Sibiryakov, \emph{{On the Extra Mode and Inconsistency of Horava Gravity}}, \href{https://doi.org/10.1088/1126-6708/2009/10/029}{\emph{JHEP} {\bfseries 10} (2009) 029} [\href{https://arxiv.org/abs/0906.3046}{{\ttfamily 0906.3046}}].

\bibitem{Cheung:2007st}
C.~Cheung, P.~Creminelli, A.~L. Fitzpatrick, J.~Kaplan and L.~Senatore, \emph{{The Effective Field Theory of Inflation}}, \href{https://doi.org/10.1088/1126-6708/2008/03/014}{\emph{JHEP} {\bfseries 03} (2008) 014} [\href{https://arxiv.org/abs/0709.0293}{{\ttfamily 0709.0293}}].

\bibitem{Gubitosi:2012hu}
G.~Gubitosi, F.~Piazza and F.~Vernizzi, \emph{{The Effective Field Theory of Dark Energy}}, \href{https://doi.org/10.1088/1475-7516/2013/02/032}{\emph{JCAP} {\bfseries 02} (2013) 032} [\href{https://arxiv.org/abs/1210.0201}{{\ttfamily 1210.0201}}].

\bibitem{Lin:2017oow}
C.~Lin and S.~Mukohyama, \emph{{A Class of Minimally Modified Gravity Theories}}, \href{https://doi.org/10.1088/1475-7516/2017/10/033}{\emph{JCAP} {\bfseries 10} (2017) 033} [\href{https://arxiv.org/abs/1708.03757}{{\ttfamily 1708.03757}}].

\bibitem{Iyonaga:2018vnu}
A.~Iyonaga, K.~Takahashi and T.~Kobayashi, \emph{{Extended Cuscuton: Formulation}}, \href{https://doi.org/10.1088/1475-7516/2018/12/002}{\emph{JCAP} {\bfseries 12} (2018) 002} [\href{https://arxiv.org/abs/1809.10935}{{\ttfamily 1809.10935}}].

\bibitem{Gao:2019twq}
X.~Gao and Z.-B. Yao, \emph{{Spatially covariant gravity theories with two tensorial degrees of freedom: the formalism}}, \href{https://doi.org/10.1103/PhysRevD.101.064018}{\emph{Phys. Rev. D} {\bfseries 101} (2020) 064018} [\href{https://arxiv.org/abs/1910.13995}{{\ttfamily 1910.13995}}].

\bibitem{Lin:2020nro}
J.~Lin, Y.~Gong, Y.~Lu and F.~Zhang, \emph{{Spatially covariant gravity with a dynamic lapse function}}, \href{https://doi.org/10.1103/PhysRevD.103.064020}{\emph{Phys. Rev. D} {\bfseries 103} (2021) 064020} [\href{https://arxiv.org/abs/2011.05739}{{\ttfamily 2011.05739}}].

\bibitem{Iyonaga:2021yfv}
A.~Iyonaga and T.~Kobayashi, \emph{{Distinguishing modified gravity with just two tensorial degrees of freedom from general relativity: Black holes, cosmology, and matter coupling}}, \href{https://doi.org/10.1103/PhysRevD.104.124020}{\emph{Phys. Rev. D} {\bfseries 104} (2021) 124020} [\href{https://arxiv.org/abs/2109.10615}{{\ttfamily 2109.10615}}].

\bibitem{Vainshtein:1972sx}
A.~I. Vainshtein, \emph{{To the problem of nonvanishing gravitation mass}}, \href{https://doi.org/10.1016/0370-2693(72)90147-5}{\emph{Phys. Lett. B} {\bfseries 39} (1972) 393}.

\bibitem{Kimura:2011dc}
R.~Kimura, T.~Kobayashi and K.~Yamamoto, \emph{{Vainshtein screening in a cosmological background in the most general second-order scalar-tensor theory}}, \href{https://doi.org/10.1103/PhysRevD.85.024023}{\emph{Phys. Rev. D} {\bfseries 85} (2012) 024023} [\href{https://arxiv.org/abs/1111.6749}{{\ttfamily 1111.6749}}].

\bibitem{Babichev:2013usa}
E.~Babichev and C.~Deffayet, \emph{{An introduction to the Vainshtein mechanism}}, \href{https://doi.org/10.1088/0264-9381/30/18/184001}{\emph{Class. Quant. Grav.} {\bfseries 30} (2013) 184001} [\href{https://arxiv.org/abs/1304.7240}{{\ttfamily 1304.7240}}].

\bibitem{Kobayashi:2014ida}
T.~Kobayashi, Y.~Watanabe and D.~Yamauchi, \emph{{Breaking of Vainshtein screening in scalar-tensor theories beyond Horndeski}}, \href{https://doi.org/10.1103/PhysRevD.91.064013}{\emph{Phys. Rev. D} {\bfseries 91} (2015) 064013} [\href{https://arxiv.org/abs/1411.4130}{{\ttfamily 1411.4130}}].

\bibitem{Crisostomi:2017lbg}
M.~Crisostomi and K.~Koyama, \emph{{Vainshtein mechanism after GW170817}}, \href{https://doi.org/10.1103/PhysRevD.97.021301}{\emph{Phys. Rev. D} {\bfseries 97} (2018) 021301} [\href{https://arxiv.org/abs/1711.06661}{{\ttfamily 1711.06661}}].

\bibitem{Langlois:2017dyl}
D.~Langlois, R.~Saito, D.~Yamauchi and K.~Noui, \emph{{Scalar-tensor theories and modified gravity in the wake of GW170817}}, \href{https://doi.org/10.1103/PhysRevD.97.061501}{\emph{Phys. Rev. D} {\bfseries 97} (2018) 061501} [\href{https://arxiv.org/abs/1711.07403}{{\ttfamily 1711.07403}}].

\bibitem{Dima:2017pwp}
A.~Dima and F.~Vernizzi, \emph{{Vainshtein Screening in Scalar-Tensor Theories before and after GW170817: Constraints on Theories beyond Horndeski}}, \href{https://doi.org/10.1103/PhysRevD.97.101302}{\emph{Phys. Rev. D} {\bfseries 97} (2018) 101302} [\href{https://arxiv.org/abs/1712.04731}{{\ttfamily 1712.04731}}].

\bibitem{Hirano:2019scf}
S.~Hirano, T.~Kobayashi and D.~Yamauchi, \emph{{Screening mechanism in degenerate higher-order scalar-tensor theories evading gravitational wave constraints}}, \href{https://doi.org/10.1103/PhysRevD.99.104073}{\emph{Phys. Rev. D} {\bfseries 99} (2019) 104073} [\href{https://arxiv.org/abs/1903.08399}{{\ttfamily 1903.08399}}].

\bibitem{Crisostomi:2019yfo}
M.~Crisostomi, M.~Lewandowski and F.~Vernizzi, \emph{{Vainshtein regime in scalar-tensor gravity: Constraints on degenerate higher-order scalar-tensor theories}}, \href{https://doi.org/10.1103/PhysRevD.100.024025}{\emph{Phys. Rev. D} {\bfseries 100} (2019) 024025} [\href{https://arxiv.org/abs/1903.11591}{{\ttfamily 1903.11591}}].

\bibitem{Kobayashi:2023lyt}
T.~Kobayashi and T.~Hiramatsu, \emph{{Weak-field regime of scalar-tensor theories with an instantaneous mode}},  \href{https://arxiv.org/abs/2310.11041}{{\ttfamily 2310.11041}}.

\bibitem{Will:2005va}
C.~M. Will, \emph{{The Confrontation between general relativity and experiment}}, \href{https://doi.org/10.12942/lrr-2006-3}{\emph{Living Rev. Rel.} {\bfseries 9} (2006) 3} [\href{https://arxiv.org/abs/gr-qc/0510072}{{\ttfamily gr-qc/0510072}}].

\bibitem{Will:2018bme}
C.~M. Will, \emph{{Theory and Experiment in Gravitational Physics}}. Cambridge University Press, 2018.

\bibitem{Jain:2010ka}
B.~Jain and J.~Khoury, \emph{{Cosmological Tests of Gravity}}, \href{https://doi.org/10.1016/j.aop.2010.04.002}{\emph{Annals Phys.} {\bfseries 325} (2010) 1479} [\href{https://arxiv.org/abs/1004.3294}{{\ttfamily 1004.3294}}].

\bibitem{Joyce:2014kja}
A.~Joyce, B.~Jain, J.~Khoury and M.~Trodden, \emph{{Beyond the Cosmological Standard Model}}, \href{https://doi.org/10.1016/j.physrep.2014.12.002}{\emph{Phys. Rept.} {\bfseries 568} (2015) 1} [\href{https://arxiv.org/abs/1407.0059}{{\ttfamily 1407.0059}}].

\bibitem{Koyama:2015vza}
K.~Koyama, \emph{{Cosmological Tests of Modified Gravity}}, \href{https://doi.org/10.1088/0034-4885/79/4/046902}{\emph{Rept. Prog. Phys.} {\bfseries 79} (2016) 046902} [\href{https://arxiv.org/abs/1504.04623}{{\ttfamily 1504.04623}}].

\bibitem{Arai:2022ilw}
S.~Arai et~al., \emph{{Cosmological gravity probes: Connecting recent theoretical developments to forthcoming observations}}, \href{https://doi.org/10.1093/ptep/ptad052}{\emph{PTEP} {\bfseries 2023} (2023) 072E01} [\href{https://arxiv.org/abs/2212.09094}{{\ttfamily 2212.09094}}].

\bibitem{Gleyzes:2013ooa}
J.~Gleyzes, D.~Langlois, F.~Piazza and F.~Vernizzi, \emph{{Essential Building Blocks of Dark Energy}}, \href{https://doi.org/10.1088/1475-7516/2013/08/025}{\emph{JCAP} {\bfseries 08} (2013) 025} [\href{https://arxiv.org/abs/1304.4840}{{\ttfamily 1304.4840}}].

\bibitem{Bellini:2014fua}
E.~Bellini and I.~Sawicki, \emph{{Maximal freedom at minimum cost: linear large-scale structure in general modifications of gravity}}, \href{https://doi.org/10.1088/1475-7516/2014/07/050}{\emph{JCAP} {\bfseries 07} (2014) 050} [\href{https://arxiv.org/abs/1404.3713}{{\ttfamily 1404.3713}}].

\bibitem{Langlois:2017mxy}
D.~Langlois, M.~Mancarella, K.~Noui and F.~Vernizzi, \emph{{Effective Description of Higher-Order Scalar-Tensor Theories}}, \href{https://doi.org/10.1088/1475-7516/2017/05/033}{\emph{JCAP} {\bfseries 05} (2017) 033} [\href{https://arxiv.org/abs/1703.03797}{{\ttfamily 1703.03797}}].

\bibitem{Lombriser:2018guo}
L.~Lombriser, \emph{{Parametrizations for tests of gravity}}, \href{https://doi.org/10.1142/S0218271818480024}{\emph{Int. J. Mod. Phys. D} {\bfseries 27} (2018) 1848002} [\href{https://arxiv.org/abs/1908.07892}{{\ttfamily 1908.07892}}].

\bibitem{Renevey:2020tvr}
C.~Renevey, J.~Kennedy and L.~Lombriser, \emph{{Parameterised post-Newtonian formalism for the effective field theory of dark energy via screened reconstructed Horndeski theories}}, \href{https://doi.org/10.1088/1475-7516/2020/12/032}{\emph{JCAP} {\bfseries 12} (2020) 032} [\href{https://arxiv.org/abs/2006.09910}{{\ttfamily 2006.09910}}].

\bibitem{Zumalacarregui:2016pph}
M.~Zumalac\'arregui, E.~Bellini, I.~Sawicki, J.~Lesgourgues and P.~G. Ferreira, \emph{{hi\_class: Horndeski in the Cosmic Linear Anisotropy Solving System}}, \href{https://doi.org/10.1088/1475-7516/2017/08/019}{\emph{JCAP} {\bfseries 08} (2017) 019} [\href{https://arxiv.org/abs/1605.06102}{{\ttfamily 1605.06102}}].

\bibitem{Hiramatsu:2020fcd}
T.~Hiramatsu and D.~Yamauchi, \emph{{Testing gravity theories with cosmic microwave background in the degenerate higher-order scalar-tensor theory}}, \href{https://doi.org/10.1103/PhysRevD.102.083525}{\emph{Phys. Rev. D} {\bfseries 102} (2020) 083525} [\href{https://arxiv.org/abs/2004.09520}{{\ttfamily 2004.09520}}].

\bibitem{Blas:2011zd}
D.~Blas and H.~Sanctuary, \emph{{Gravitational Radiation in Ho\v{r}ava Gravity}}, \href{https://doi.org/10.1103/PhysRevD.84.064004}{\emph{Phys. Rev. D} {\bfseries 84} (2011) 064004} [\href{https://arxiv.org/abs/1105.5149}{{\ttfamily 1105.5149}}].

\bibitem{Lin:2013tua}
K.~Lin, S.~Mukohyama, A.~Wang and T.~Zhu, \emph{{Post-Newtonian approximations in the Ho\v{r}ava-Lifshitz gravity with extra U(1) symmetry}}, \href{https://doi.org/10.1103/PhysRevD.89.084022}{\emph{Phys. Rev. D} {\bfseries 89} (2014) 084022} [\href{https://arxiv.org/abs/1310.6666}{{\ttfamily 1310.6666}}].

\bibitem{Qiao:2021fwi}
J.~Qiao, T.~Zhu, G.~Li and W.~Zhao, \emph{{Post-Newtonian parameters of ghost-free parity-violating gravities}}, \href{https://doi.org/10.1088/1475-7516/2022/04/054}{\emph{JCAP} {\bfseries 04} (2022) 054} [\href{https://arxiv.org/abs/2110.09033}{{\ttfamily 2110.09033}}].

\bibitem{BeltranJimenez:2015sgd}
J.~Beltran~Jimenez, F.~Piazza and H.~Velten, \emph{{Evading the Vainshtein Mechanism with Anomalous Gravitational Wave Speed: Constraints on Modified Gravity from Binary Pulsars}}, \href{https://doi.org/10.1103/PhysRevLett.116.061101}{\emph{Phys. Rev. Lett.} {\bfseries 116} (2016) 061101} [\href{https://arxiv.org/abs/1507.05047}{{\ttfamily 1507.05047}}].

\bibitem{LIGOScientific:2017vwq}
{\scshape LIGO Scientific, Virgo} collaboration, B.~P. Abbott et~al., \emph{{GW170817: Observation of Gravitational Waves from a Binary Neutron Star Inspiral}}, \href{https://doi.org/10.1103/PhysRevLett.119.161101}{\emph{Phys. Rev. Lett.} {\bfseries 119} (2017) 161101} [\href{https://arxiv.org/abs/1710.05832}{{\ttfamily 1710.05832}}].

\bibitem{LIGOScientific:2017zic}
{\scshape LIGO Scientific, Virgo, Fermi-GBM, INTEGRAL} collaboration, B.~P. Abbott et~al., \emph{{Gravitational Waves and Gamma-rays from a Binary Neutron Star Merger: GW170817 and GRB 170817A}}, \href{https://doi.org/10.3847/2041-8213/aa920c}{\emph{Astrophys. J. Lett.} {\bfseries 848} (2017) L13} [\href{https://arxiv.org/abs/1710.05834}{{\ttfamily 1710.05834}}].

\bibitem{Bertotti:2003rm}
B.~Bertotti, L.~Iess and P.~Tortora, \emph{{A test of general relativity using radio links with the Cassini spacecraft}}, \href{https://doi.org/10.1038/nature01997}{\emph{Nature} {\bfseries 425} (2003) 374}.

\bibitem{Verma:2013ata}
A.~Verma, A.~Fienga, J.~Laskar, H.~Manche and M.~Gastineau, \emph{{Use of MESSENGER radioscience data to improve planetary ephemeris and to test general relativity}}, \href{https://doi.org/10.1051/0004-6361/201322124}{\emph{Astron. Astrophys.} {\bfseries 561} (2014) A115} [\href{https://arxiv.org/abs/1306.5569}{{\ttfamily 1306.5569}}].

\bibitem{Shao:2012eg}
L.~Shao and N.~Wex, \emph{{New tests of local Lorentz invariance of gravity with small-eccentricity binary pulsars}}, \href{https://doi.org/10.1088/0264-9381/29/21/215018}{\emph{Class. Quant. Grav.} {\bfseries 29} (2012) 215018} [\href{https://arxiv.org/abs/1209.4503}{{\ttfamily 1209.4503}}].

\bibitem{Shao:2013wga}
L.~Shao, R.~N. Caballero, M.~Kramer, N.~Wex, D.~J. Champion and A.~Jessner, \emph{{A new limit on local Lorentz invariance violation of gravity from solitary pulsars}}, \href{https://doi.org/10.1088/0264-9381/30/16/165019}{\emph{Class. Quant. Grav.} {\bfseries 30} (2013) 165019} [\href{https://arxiv.org/abs/1307.2552}{{\ttfamily 1307.2552}}].

\bibitem{Blas:2010hb}
D.~Blas, O.~Pujolas and S.~Sibiryakov, \emph{{Models of non-relativistic quantum gravity: The Good, the bad and the healthy}}, \href{https://doi.org/10.1007/JHEP04(2011)018}{\emph{JHEP} {\bfseries 04} (2011) 018} [\href{https://arxiv.org/abs/1007.3503}{{\ttfamily 1007.3503}}].

\bibitem{Afshordi:2006ad}
N.~Afshordi, D.~J.~H. Chung and G.~Geshnizjani, \emph{{Cuscuton: A Causal Field Theory with an Infinite Speed of Sound}}, \href{https://doi.org/10.1103/PhysRevD.75.083513}{\emph{Phys. Rev. D} {\bfseries 75} (2007) 083513} [\href{https://arxiv.org/abs/hep-th/0609150}{{\ttfamily hep-th/0609150}}].

\bibitem{Hu:2021yaq}
Y.-M. Hu and X.~Gao, \emph{{Spatially covariant gravity with 2 degrees of freedom: Perturbative analysis}}, \href{https://doi.org/10.1103/PhysRevD.104.104007}{\emph{Phys. Rev. D} {\bfseries 104} (2021) 104007} [\href{https://arxiv.org/abs/2104.07615}{{\ttfamily 2104.07615}}].

\bibitem{Franzin:2023rdl}
E.~Franzin, S.~Liberati and J.~Mazza, \emph{{A Kerr Black Hole in Einstein--\AE{}ther Gravity}},  \href{https://arxiv.org/abs/2312.06891}{{\ttfamily 2312.06891}}.

\bibitem{Gurses:2024jkx}
M.~Gurses, C.~Senturk and B.~Tekin, \emph{{Minimal Einstein-Aether Theory}},  \href{https://arxiv.org/abs/2402.07068}{{\ttfamily 2402.07068}}.

\end{thebibliography}\endgroup
\bibliographystyle{JHEP}
\end{document}